\documentclass[a4paper,11pt]{article}
\usepackage[utf8]{inputenc}
\usepackage{jcappub} 
\usepackage{lineno}
\usepackage{siunitx}
\usepackage{booktabs}
\usepackage{subcaption}
\usepackage[T1]{fontenc} 
\usepackage[dvipsnames]{xcolor}

\title{Forecasting a Cosmological 21 cm Stacking Detection with Early HIRAX Data}

\author{Pascal Hitz,}
\author{Devin Crichton,}
\author{Alexandre Refregier}
\affiliation{Institute for Particle Physics and Astrophysics, ETH Zurich, 8093 Zurich, Switzerland}

\emailAdd{hitzpa@phys.ethz.ch}

\abstract{Post-reionization \qty{21}{\cm} intensity mapping provides a powerful technique for efficiently mapping the Universe on large scales. The Hydrogen Intensity and Real-time Analysis eXperiment (HIRAX) is a compact and highly redundant radio interferometer designed to exploit this technique. We present a forward modelling pipeline for forecasting a statistical detection of the cosmological \qty{21}{\cm} signal with HIRAX through cross-correlation with an external source catalogue. We simulate the cosmological signal using the halo model and propagate it, together with Galactic and extragalactic foreground emission and instrumental noise, through a model of an early-stage 36-dish HIRAX array configuration. We generate simulated visibilities, apply delay-domain foreground filtering, and construct cross-correlation-ready ringmaps for an observing time of 120 days with a single zenith pointing over the frequency range from \qtyrange{700}{800}{\MHz}, covering approximately \qty{3547}{\deg^2} of the sky. We investigate the prospects for an early statistical signal detection by stacking these ringmaps at the known three-dimensional positions of sources from an overlapping external survey. Specifically, we quantify how delay-domain foreground filtering, the number of stacking sources, catalogue completeness, and source redshift uncertainties affect the detection significance. We find that redshift precision is the most critical requirement: while spectroscopic redshifts with $\sigma_z / (1+z) \sim 0.001$ preserve a significant stacking signal even under realistic foreground filtering and substantial catalogue incompleteness, photometric redshift uncertainties of $\sigma_z / (1+z) \sim 0.01$ strongly suppress the signal for realistic delay cutoffs. Applying the framework to a Euclid-like spectroscopic galaxy sample, we recover the signal across all considered delay cutoffs, achieving a minimum signal-to-noise ratio of 11 for our most conservative cut of \qty{100}{\ns}. This work provides the groundwork for a stacking analysis data pipeline. Our results demonstrate that stacking with spectroscopic surveys is a promising avenue for an early detection of the cosmological \qty{21}{\cm} signal with initial stages of HIRAX.}

\begin{document}
\maketitle
\flushbottom

\section{Introduction}
\label{sec:Introduction}
Measuring the statistical distribution of large-scale structure in the Universe provides a powerful probe to constrain the cosmological model that describes its evolution, energy components, as well as the growth of structure. To this end, large galaxy surveys have significantly contributed to mapping the three-dimensional distribution of matter over large cosmological volumes up to redshifts $z \sim 1$ \cite{2dFGRS_2005, BOSS_2017, eBOSS_2021, WiggleZ_2012, DESI_2025}. At higher redshifts, however, it becomes increasingly difficult to detect sufficiently large numbers of galaxies and determine their redshifts well enough to infer their radial positions. A complementary approach that can efficiently map larger cosmological volumes is offered by \qty{21}{\cm} intensity mapping \cite{Battye_2004, McQuinn_2006, Wyithe_Loeb_2007, Chang_2008, Loeb_Wyithe_2008, Mao_2008, Pritchard_Loeb_2008, Peterson_2009, Seo_2010, Bagla_2010, Lidz_2011, Ansari_2012, Bull_2015}. \\
Neutral hydrogen (HI) emits radiation at a rest-frame wavelength of approximately \qty{21}{\cm}, corresponding to a frequency of $\nu_{21} \simeq \qty{1420.406}{\MHz}$, through the hyperfine transition between the aligned and anti-aligned spin configurations of the proton and electron. The abundance of HI throughout the expansion history of the Universe provides unique access to a broad range of cosmic epochs, extending from the post-reionization Universe to the Epoch of Reionization, Cosmic Dawn, and potentially the Dark Ages before the first luminous sources formed. After the Epoch of Reionization ($z \lesssim 6$), most of the remaining HI is expected to reside in the dense self-shielded regions within galaxies, acting as a biased tracer of the underlying dark matter field similarly to galaxies \cite{Villaescusa-Navarro_2018}. However, the well-defined rest frequency of the \qty{21}{\cm} transition allows the observed frequency to be directly mapped to redshift. It is therefore not necessary to resolve individual galaxies to obtain radial information, which can otherwise be observationally cumbersome. Instead, in \qty{21}{\cm} intensity mapping the integrated emission of unresolved HI is measured, thereby enabling the three-dimensional matter distribution to be traced efficiently over vast cosmological volumes.  \\
The growing utility of this technique is reflected in the increasing number of instruments, either already in operation or under development, that are designed for or capable of performing \qty{21}{\cm} intensity mapping measurements. These target both the post-reionization era (e.g. SKA-MID \cite{Advancing_astrophysics_skaii_2026, Cosmology_HI_Intensity_Mapping_SKA_2026, SKA_MID_2026}, MeerKAT \cite{MeerKAT_2016}, BINGO \cite{BINGO_2019}, FAST \cite{FAST_2011}, GBT \cite{GBT_2009}, Parkes \cite{Parkes_1996}, CHIME \cite{CHIME_2014}, CHORD \cite{CHORD_2019}, HIRAX \cite{HIRAX_2016, HIRAX_2022}, Tianlai \cite{Tianlai_2015}, and PUMA \cite{PUMA_2019}) and the Epoch of Reionization (e.g. SKA-LOW \cite{Advancing_astrophysics_skaii_2026, Cosmology_HI_Intensity_Mapping_SKA_2026, SKA_LOW_2026}, HERA \cite{HERA_2017}, LOFAR \cite{LOFAR_2013}, and MWA \cite{MWA_2013}). Several detections of the post-reionization \qty{21}{\cm} signal with intensity mapping have already been reported through source stacking \cite{Parkes_stacking_2dFGRS_2020, CHIME_stacking_eBOSS_2023, MeerKAT_stacking_GAMA_2025}, other cross-correlations \cite{HIPASS_cross_6dFGRS_2009, GBT_cross_DEEP2_2010, GBT_cross_WiggleZ_2013, Parkes_cross_2dFGRS_2018, Parkes_cross_WiggleZ_2021, GBT_cross_eBOSS_WiggleZ_2022, MeerKAT_cross_WiggleZ_2023, CHIME_cross_eBOSSLyalpha_2024, MeerKAT_cross_WiggleZ_2025, MeerKAT_cross_GAMA_2025}, and, more recently, auto-correlation \cite{CHIME_auto_2026, MeerKAT_auto_2026, MeerKLASS_auto_2026}. \\
HIRAX, the Hydrogen Intensity and Real-time Analysis eXperiment, is a radio interferometer currently under construction in the Karoo desert of South Africa. Its primary science goals include measuring baryon acoustic oscillations and constraining the properties of dark energy, while also enabling the search for and study of pulsars and radio transients, including fast radio bursts. In its final configuration, HIRAX is planned to consist of 1024 individual \qty{6}{\m} dishes arranged in a compact, highly redundant array, observing the southern sky over a frequency range of \qtyrange{400}{800}{\MHz} with a spectral resolution of \qty{0.39}{\MHz}. This frequency range corresponds to redshifts $0.78 < z < 2.55$ for the \qty{21}{\cm} line. HIRAX is designed to operate as a drift-scan telescope, with the dishes fixed at a given declination while the rotation of the Earth sweeps the field of view across the sky. By periodically repointing the array in declination, the survey is designed to cover a substantial fraction of the southern sky, spanning declinations from \qtyrange{-60}{0}{\degree}. Its location, observing strategy, and instrumental design make HIRAX highly complementary to CHIME, which primarily surveys the northern sky. \\
In preparation for the analysis and interpretation of real HIRAX data, realistic simulations of both the sky signal and the instrument response play an important role. In a previous paper \cite{Hitz_2025}, we presented and validated simulations of the post-reionization cosmological HI signal based on a halo-model approach. In this work, we use these simulations, supplemented with foreground emission and instrumental noise, as an input sky signal to forward model realistic HIRAX observations through to ringmaps to be used for cross-correlation. As the HIRAX array is being deployed in stages, with commissioning and verification periods between successive phases, a central goal of this paper is to investigate whether a statistically significant HI signal can already be detected with a preliminary array configuration and limited coverage of the full survey volume. Given the comparatively weak cosmological HI signal relative to the much brighter Galactic and extragalactic foregrounds and instrumental noise across the observed frequency range, we focus on a cross-correlation detection, which can mitigate the impact of HIRAX-specific instrumental systematics. In particular, we investigate source stacking as a simple and direct cross-correlation technique, in which the \qty{21}{\cm} signal in the final maps is averaged at the known three-dimensional positions of emitting sources identified by an overlapping external survey. \\

The rest of the paper is organised as follows: In section \ref{sec:Simulated HI Sky Signal}, we briefly review the simulation of the cosmological HI sky signal used in this work. In section \ref{sec:Forward-Modelled HIRAX Observation}, we describe the forward modelling pipeline used to simulate HIRAX observations, including the preliminary instrument model that we adopt here, the visibility generation, foreground filtering, and the final map construction. Section \ref{sec:Source Stacking} introduces the stacking analysis used to extract the HI signal and investigates the requirements on the source catalogue and foreground filtering needed for a statistically significant detection. In section \ref{sec:Results}, we present the resulting detection forecasts and apply our analysis to a realistic cross-correlation scenario with the Euclid spectroscopic galaxy sample. Finally, we summarise our results and conclusions in section \ref{sec:Conclusion}. \\
Throughout this work, whenever a cosmology is required, we adopt the same flat $\Lambda$CDM cosmology as in our previous work \cite{Hitz_2025}, with cosmological parameters based on the Planck 2018 results \cite{Planck_2018}: $\Omega_\text{m} = 0.3153$, $\Omega_{\Lambda} = 0.6847$, $\Omega_\text{b} = 0.0493$, $h = 0.6736$, $\sigma_8 = 0.8111$, $n_\text{s} = 0.9649$, and $Y_\text{p} = 0.24$.

\section{Simulated HI Sky Signal}
\label{sec:Simulated HI Sky Signal}
As the input cosmological HI sky signal, we use simulated tomographic HI brightness temperature maps. The production and validation of these maps are described in our previous paper \cite{Hitz_2025}, which we briefly summarise here.\footnote{The simulation dataset is available at \url{https://cosmology.ethz.ch/research/software-lab/data-products/cosmological-neutral-hydrogen-simulation.html}.} A detailed comparison between this simulation and other cosmological HI simulations in the literature is also provided in \cite{Ronconi_2026}. \\

The HI simulations are based on the \texttt{PINOCCHIO} code \cite{Monaco_2002}, which is used to generate the past light cone of dark matter halos. By combining Lagrangian Perturbation Theory with the Extended Press \& Schechter formalism for non-spherical collapse \cite{Bond_1991, Monaco_1995, Monaco_1997} to generate the dark matter halo light cone on-the-fly, \texttt{PINOCCHIO} provides a computationally efficient approximate description of structure formation. Despite its approximate nature, it reproduces the halo clustering statistics of full \emph{N}-body simulations to within 10\%, which is sufficient for forecasting purposes. The dark matter halos are subsequently populated with HI according to the empirical HI-halo mass relation \cite{Padmanabhan_2017}:
\begin{equation}
M_\text{HI}(M,z) = \alpha f_\text{H,c} M \biggl (\frac{M}{10^{11} \, h^{-1} \text{M}_{\odot}} \biggr)^{\beta} \exp \biggl[- \Bigl( \frac{v_\text{c,0}}{v_\text{c}(M,z)} \Bigr)^3 \biggr].
\label{eq:MHI-M relation}
\end{equation}
In this expression $M_\text{HI}$ is the HI mass assigned to a halo with mass $M$, $\alpha$ denotes the neutral fraction of HI in the halo relative to the cosmic hydrogen fraction $f_\text{H,c} = (1-Y_\text{p})\Omega_\text{b} / \Omega_\text{m}$, and $v_\text{c}$ is the virial velocity of the host halo. The free parameters are set to their best-fit values, $\alpha = 0.09$, $\beta = -0.58$, and $\log_{10}[ v_\text{c,0}/(\mathrm{km \, s}^{-1})] = 1.56$. Through the dependence of the virial velocity on halo mass \citep{Barnes_2014}, the exponential term imposes a low-mass cutoff, reflecting the reduced ability of these halos to self-shield HI against ionizing radiation. \\
Although individual low-mass halos contain only small amounts of HI, their large abundance makes them important contributors to the total cosmic HI content. To achieve the required mass resolution, the simulated volume is restricted to a light cone covering declinations from \qtyrange{-15}{-45}{\degree} and frequencies from \qtyrange{700}{800}{\MHz}, i.e. redshifts between 0.77 and 1.03. In our simulations, we adopt a slightly extended redshift range down to $z=0.77$, instead of $z=0.78$, to ensure full coverage of the HIRAX frequency band. Running \texttt{PINOCCHIO} with a box size of \qty{1}{h^{-1} Gpc}, $6700^3$ simulation particles per box, and a minimum halo size of 10 particles yields a halo mass resolution of \qty{4.3e9}{M_{\odot}}, recovering approximately 98\% of the total HI content.\\
Finally, the HI masses are converted into brightness temperatures using the relation presented in \cite{Bull_2015}, and the resulting signal is pixelised into HEALPix \cite{HEALPix_2005} maps. In \cite{Hitz_2025}, maps with a bandwidth of \qty{5}{\MHz} and an \emph{NSIDE} parameter of 512 were analysed, such that the maps are dominated by the cosmological signal rather than Poisson noise when compared with theoretical predictions. For our purposes here, we instead choose a higher frequency resolution, producing 64 tomographic maps with a width of \qty{1.5625}{\MHz}, while keeping the same \emph{NSIDE} parameter. \\
The simulation was validated in \cite{Hitz_2025} against an analytical dark matter and HI halo model implemented in \texttt{PyCosmo} \cite{Refregier_2018, Tarsitano_2021, Moser_2022, Hitz_2025}, a Python package for theoretical cosmological predictions. Agreement at the level of a few percent was found for the halo mass function and dark matter and HI mass densities, while the three-dimensional and angular power spectra were reproduced to within 10\%.

\section{Forward-Modelled HIRAX Observation}
\label{sec:Forward-Modelled HIRAX Observation}
In this section, we describe the pipeline for performing mock HIRAX observations of the simulated HI sky signal presented in section \ref{sec:Simulated HI Sky Signal}. We closely follow the CHIME pipeline described in \cite{CHIME_stacking_eBOSS_2023} to produce the so-called \emph{ringmaps}. In the process of simulating the operation of the HIRAX instrument, we make extensive use of the Python packages \texttt{cora}, \texttt{caput}, \texttt{driftscan}, and \texttt{draco} \cite{cora_2024, caput_2024, driftscan_2026, draco_2024}. These were adapted where necessary to support the HIRAX array configuration. They are based on the \emph{m}-mode formalism \cite{Shaw_2014, Shaw_2015}, for which we provide a brief overview here before summarising the production of the mock data.

\subsection{\emph{m}-mode Formalism}
\label{sec:m-mode Formalism}
The fundamental observables of an interferometer such as HIRAX are the visibilities, the cross-correlations of the signals measured by pairs of feeds, denoted by \emph{i} and \emph{j}. HIRAX will be operating in drift-scan mode, continuously observing the sky in different frequency channels at a fixed declination, with Earth's rotation sweeping out a stripe across the sky. Consequently, the visibilities can be written as a function of frequency $\nu$ and rotation angle $\phi$:
\begin{equation}
    \begin{split}
        V_{ij}(\nu,\phi)
        &=
        \int
        \Bigl[
        B_{ij}^{T}(\hat{n};\nu,\phi)\,T(\hat{n};\nu)
        +
        B_{ij}^{Q}(\hat{n};\nu,\phi)\,Q(\hat{n};\nu)
        \\
        &\qquad
        +
        B_{ij}^{U}(\hat{n};\nu,\phi)\,U(\hat{n};\nu)
        +
        B_{ij}^{V}(\hat{n};\nu,\phi)\,V(\hat{n};\nu)
        \Bigr]
        \,d^{2}\hat{n}
        +
        n_{ij}(\nu,\phi),
    \end{split}
\label{eq:visibility}
\end{equation}
where the integral extends over all sky directions $\hat{n}$. The sky signal is described by the Stokes parameters \emph{T}, \emph{Q}, \emph{U}, and \emph{V}, while the beam transfer functions $B^{X}_{ij}$ characterise the instrument model, and $n_{ij}$ the noise. Since we only consider unpolarised HI emission, we neglect the polarised components in the following. However, their treatment is analogous to that of the intensity \emph{T} \cite{Shaw_2015}. Expanding both $B^{T}_{ij}$ and \emph{T} in spherical harmonics transforms the sky integral into a sum over spherical harmonic modes:
\begin{equation}
    V_{ij}(\nu, \phi)
    =
    \sum_{lm}
    B^{T}_{ij;lm}(\nu, \phi)\,a^{T}_{lm}(\nu)
    + n_{ij}(\nu, \phi),
\label{eq:visibility_harmonic}
\end{equation}
where $a^{T}_{lm}$ are the spherical harmonic coefficients of \emph{T} and $B^{T}_{ij;lm}$ are the complex conjugates of the corresponding spherical harmonic coefficients of the beam transfer function. Since the visibilities are periodic functions of the Earth's rotation angle, they can be Fourier transformed with respect to $\phi$, isolating each azimuthal mode \emph{m}, yielding the so-called \emph{m}-modes:
\begin{equation}
    V_{ij;m}(\nu)
    =
    \int \frac{d\phi}{2\pi}\,
    V_{ij}(\nu,\phi)\,e^{-im\phi}
    =
    \sum_{l}
    B^{T}_{ij;lm}(\nu)\,a^{T}_{lm}(\nu)
    +
    n_{ij;m}(\nu),
\label{eq:m-modes}
\end{equation}
with the beam transfer matrices $B^{T}_{ij;lm}(\nu) = B^{T}_{ij;lm}(\nu,\phi=0)$. \\
We have introduced $\phi$ as the Earth's rotation angle. However, in our idealised simulation, we neglect the small differences between this angle and celestial right ascension and identify $\phi$ with right ascension throughout the rest of this work.

\subsection{Instrument Model}
\label{sec:Instrument Model}
In the first step of the mock observation pipeline, the instrument model, i.e. the beam transfer matrices $B^{T}_{ij;lm}$, is computed using \texttt{driftscan} \cite{driftscan_2026}. Since the primary objective of this work is to determine whether a preliminary HIRAX array configuration is capable of achieving a first detection of the cosmological HI signal, we consider a scaled-down version of the full array. The configuration that we consider here consists of 36 dishes with a diameter of $D=\qty{6}{\m}$ arranged in a redundant $6 \times 6$ grid. The dishes are separated by $d_x = \qty{6.5}{\m}$ in the east--west and $d_y = \qty{8.5}{\m}$ in the north--south direction. They are equipped with dual-polarisation feeds, oriented east--west and north--south. We assume a Gaussian primary beam with a full width at half maximum of $\lambda/ D$, where $\lambda$ denotes the observing wavelength. The use of a simplified beam model is partially compensated for by adopting a more aggressive foreground filtering strategy, as discussed later. The array is located at a latitude of \qty{-30.7}{\degree} and a longitude of \qty{21.5}{\degree}, corresponding to the planned HIRAX site. Instrumental noise is modelled as Gaussian, with a standard deviation proportional to the system temperature, assumed to be \qty{50}{\K}, and inversely proportional to the square root of the observing time, assumed to be 120 days, representing an expected season of HIRAX data at a fixed pointing. Given this setup, a single zenith pointing operated in drift-scan mode is simulated.

\subsection{Visibilities}
\label{sec:Visibilities}
In the next step, we use \texttt{draco} \cite{draco_2024} to generate a sidereal time stream of visibilities measured from the simulated sky. Besides the cosmological HI signal, we also consider astrophysical foregrounds that are several orders of magnitude brighter than the HI signal. The foreground emission includes Galactic synchrotron radiation and continuum radio point sources, both generated using \texttt{cora} \cite{cora_2024}. The visibilities are generated by expanding the sky signal in spherical harmonics, applying the beam transfer matrices $B^{T}_{ij;lm}$ to the $a^{T}_{lm}$ as described in eq. \eqref{eq:m-modes}, and finally performing an inverse Fourier transform over \emph{m} to get the sidereal visibilities as defined in eq. \eqref{eq:visibility}. \\
We apply foreground filtering to the visibilities before map making using the \texttt{DAYENU} method \cite{DAYENU_2021}. \texttt{DAYENU} is a delay-domain high-pass filtering technique constructed to remain effective in the presence of finite instrumental bandwidth and missing frequency channels. Although these effects are not essential in the idealised simulations considered in this work, they will be relevant for real data and we therefore incorporate the filter into the pipeline presented here. The method is based on the spectral smoothness of the foregrounds. In delay space, where the delay $\tau$ is the Fourier-conjugate to frequency $\nu$, the foregrounds are expected to be concentrated at low delays and can thus be suppressed by filtering out these modes. In this analysis, we apply a uniform delay cutoff, $\tau_\mathrm{cut}$, to all baselines, which can be regarded as a conservative filtering approach.

\subsection{Ringmaps}
\label{sec:Ringmaps}
\begin{figure}
    \centering
    \includegraphics[width=1\linewidth]{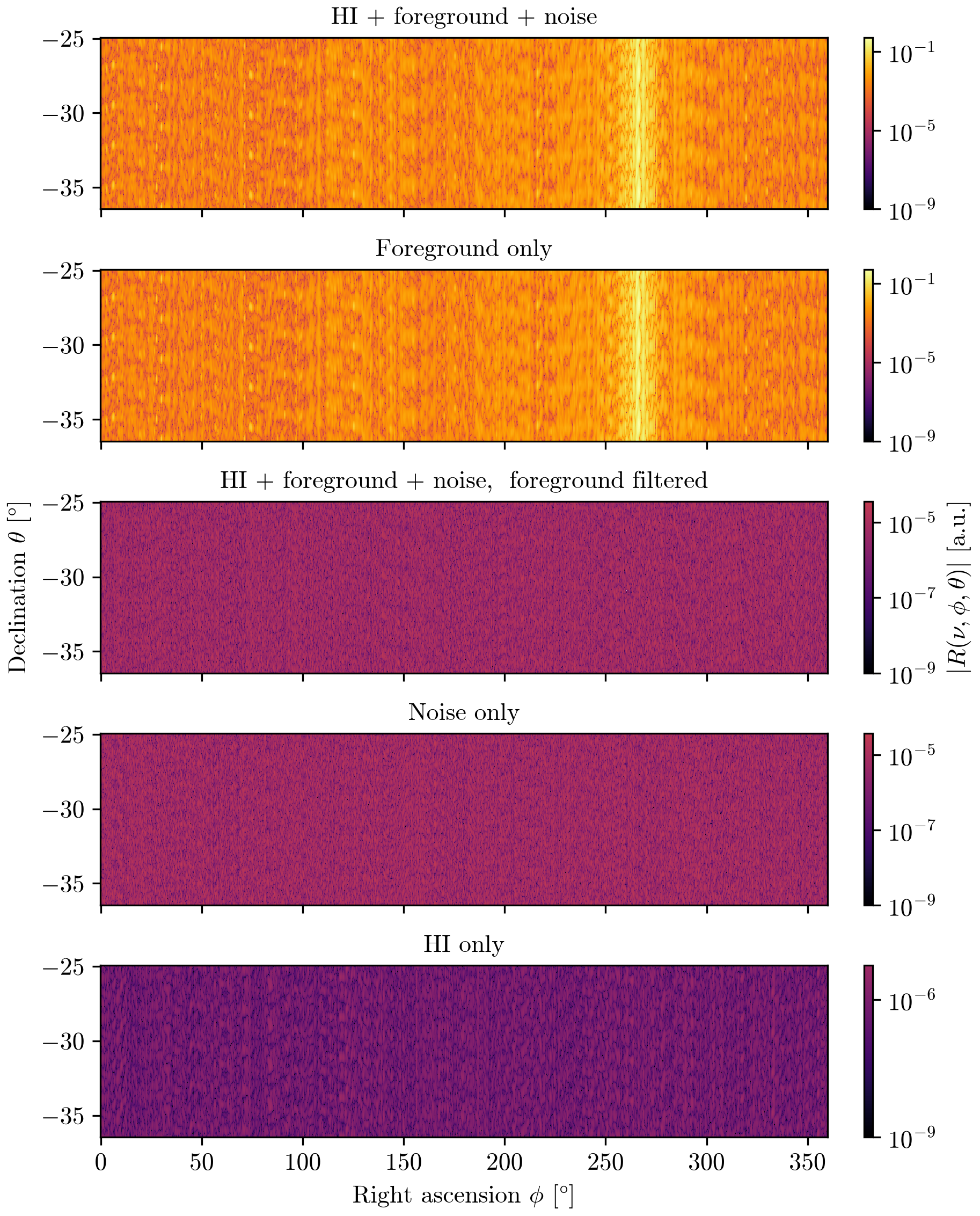}
    \caption{Absolute values of the ringmaps, $|R(\nu,\phi,\theta)|$, for the central frequency bin from \qtyrange{750}{751.5625}{\MHz}. From top to bottom: the total observed signal (HI, foregrounds, and instrumental noise), the foreground contribution alone, the total signal after foreground filtering with a delay cutoff of $\tau_\mathrm{cut}=\qty{5}{\ns}$, and the individual noise and HI components. The pronounced ringing in the north--south direction results from the lack of short north--south baselines in our array configuration.}
    \label{fig:ringmaps}
\end{figure}
The resulting sidereal visibilities are converted into a ringmap using the \texttt{RingMapMaker} implemented in \texttt{draco}. We briefly summarise the map-making procedure here and refer the reader to \cite{CHIME_stacking_eBOSS_2023} for more details. \\
The procedure is carried out in three steps. First, the redundant visibilities $V_{ij}(\nu,\phi)$ are rearranged onto a non-redundant regular Cartesian baseline grid, keeping track of the baseline redundancy needed for the weighting in the subsequent beamforming steps. The baselines between two feeds can be expressed as $\mathbf{b} = x d_x \mathbf{\hat{x}} + y d_y \mathbf{\hat{y}}$, where $\mathbf{\hat{x}}$ and $\mathbf{\hat{y}}$ are unit vectors in the east--west and north--south directions, respectively, and $x,y \in [-5,5]$ define the grid indices. In this representation, the measured visibilities $V_{ij}(\nu,\phi)$ are expressed as gridded baseline visibilities $V_{xy}(\nu,\phi)$. \\
In a second step, the gridded visibilities are beamformed in the north--south direction. For each baseline with the same east--west component \emph{x}, the visibilities are summed over the north--south baseline axis \emph{y} with the appropriate geometric phase factors, so that the signal from a specific declination $\theta$ adds coherently. This forms the so-called hybrid beamformed visibilities:
\begin{equation}
    V_x(\nu,\phi,\theta)
    =
    \sum\limits_{y} W_{xy}(\nu,\phi)
    V_{xy}(\nu,\phi)
    e^{-i 2 \pi \nu y d_y \sin(\theta -\theta_{\mathrm{tel}})/c}
\label{eq:NS beamforming}
\end{equation}
where $c$ denotes the speed of light and
\begin{equation}
    W_{xy}
    = 
    w_{xy}(\nu, \phi) / \sum\limits_{y'} w_{xy'}(\nu, \phi)
\label{eq:NS beamforming weights}
\end{equation}
are the normalised baseline weights. In this work, we use natural weighting, such that $w_{xy}(\nu, \phi)$ is given by the baseline redundancy, while auto-correlations are excluded. We calculate eq. \eqref{eq:NS beamforming} for 256 declinations equally spaced between $\pm 0.1$ in $\sin(\theta-\theta_{\mathrm{tel}})$, corresponding to a narrow stripe of $\pm \qty{5.7}{\degree}$ around zenith at $\theta_{\mathrm{tel}} = \qty{-30.7}{\degree}$ covering \qty{3547}{\deg^2} of the sky for a full rotation of the Earth.\\
In the final step, the hybrid visibilities \(V_x(\nu,\phi,\theta)\) are beamformed in the east--west direction to obtain the final ringmap $R(\nu,\phi,\theta)$. Since Earth's rotation provides the right-ascension coverage in drift-scan observations, we only consider a single central beam, centred on the local meridian. For this beam, the phase factors are unity, so beamforming reduces to a weighted sum of the hybrid beamformed visibilities over all east--west baselines \emph{x},
\begin{equation}
    R(\nu,\phi,\theta)
    =
    \sum\limits_x W_x^{\mathrm{EW}} V_x(\nu,\phi,\theta),
\label{eq:EW beamforming}
\end{equation}
with the normalised baseline weights
\begin{equation}
    W_x^{\mathrm{EW}}
    = w_{x}^{\mathrm{EW}} / \sum_{x'}w_{x'}^{\mathrm{EW}},
    \label{eq:EW beamforming weights}
\end{equation} 
where $w_{x}^{\mathrm{EW}}$ is the redundancy of the corresponding east--west baseline. The \texttt{RingMapMaker} in \texttt{draco} outputs ringmaps for the correlations between the two feed dipole polarisations, in the east--west (labelled \emph{X}) and north--south (labelled Y) directions, namely $XX$, $\mathrm{Re}[XY]$, $\mathrm{Im}[XY]$, and $YY$. To obtain the intensity, we form the average $(XX+YY)/2$. \\
Figure \ref{fig:ringmaps} presents the absolute values of example ringmaps for the central frequency bin. The first row shows the total observed signal before foreground filtering, comprising HI emission, astrophysical foregrounds, and instrumental noise. The second row shows the foreground contribution alone, while the third row shows the total signal after foreground filtering with a delay cutoff of $\tau_\mathrm{cut}=\qty{5}{\ns}$. The remaining two rows show the instrumental noise and HI components separately. Due to the lack of short north--south baselines in our array configuration, the ringmaps exhibit more pronounced ringing in the north--south direction than, for example, those obtained with CHIME \cite{CHIME_stacking_eBOSS_2023}. As expected, the unfiltered maps are strongly dominated by foreground emission, with particularly prominent emission from the Galactic plane at right ascensions around \qty{266}{\degree}. Foreground filtering substantially suppresses this emission, but the resulting maps remain dominated by instrumental noise. Consequently, the HI signal cannot be identified by visual inspection of these maps alone.

\section{Source Stacking}
\label{sec:Source Stacking}
As the HI signal from individual sources is too weak to be distinguished from residual foreground emission and instrumental noise, an initial detection will likely rely on cross-correlation with an external tracer. One particularly straightforward approach is source stacking, in which the signal in the ringmaps is co-added at the three-dimensional positions of known sources, for example, galaxies identified in overlapping surveys, thereby increasing the signal-to-noise ratio of the stacked signal. Post-reionization \qty{21}{\cm} intensity mapping detections through stacking have already been reported using Parkes, CHIME, and MeerKAT data, stacked at the positions of galaxies from 2dFGRS \cite{2dFGRS_2005}, eBOSS \cite{eBOSS_2021}, and GAMA \cite{GAMA_2015}, respectively \cite{Parkes_stacking_2dFGRS_2020, CHIME_stacking_eBOSS_2023, MeerKAT_stacking_GAMA_2025}. In the following sections, we describe our stacking procedure in detail.

\subsection{Stacked Spectral Cube}
\label{sec:Stacked Spectral Cube}
For the sake of simplicity, we use the \texttt{PINOCCHIO} halo light cone catalogue from which the simulated HI maps were generated as the source catalogue for our signal forecasts. For each halo, we identify the nearest frequency channel and map pixel in right ascension and declination. We then extract a spectral cube from the ringmaps centred on this position, spanning $\Delta\nu = \pm 10$ frequency channels, $\Delta\phi = \pm 5$ pixels in right ascension, and $\Delta\theta = \pm 30$ pixels in declination. This corresponds to approximately $\pm \qty{15.625}{\MHz}$ in frequency and $\pm \qty{1.4}{\degree}$ in $\phi$ and $\theta$. \\
Since the sensitivity to astronomical sources varies with declination according to the primary-beam response, while the thermal noise is approximately uniform, sources at different declinations contribute with different signal-to-noise ratios. An optimally weighted stacking procedure therefore weights each source according to the primary-beam response at its position. We implement this weighting by multiplying the ringmaps by the frequency-dependent Gaussian primary beam centred on zenith, $R'(\nu,\phi,\theta) = B(\nu, \theta)R(\nu,\phi,\theta)$, before performing the uniformly weighted stacking. This is equivalent, up to an overall normalisation, to applying primary-beam weights directly during stacking. \\
For each source, the corresponding spectral cube is extracted from $R'$. The extracted cubes are then stacked and averaged voxel by voxel, masking voxels that fall outside the simulated ringmaps in either frequency or declination, while right ascension is wrapped periodically. Since the HI signal associated with the catalogue objects is expected to be centred in the extracted cubes, while the instrumental noise is assumed to be random, stacking should suppress the noise and reveal a centrally peaked average HI signal. A compact way to write the stacked spectral cube is:
\begin{equation}
    s(\Delta\nu, \Delta\phi, \Delta\theta)
    =
    \frac{
    \sum_{n=1}^{N_{\mathrm{s}}}
    M(\nu_n + \Delta\nu, \, \phi_n + \Delta\phi, \, \theta_n + \Delta\theta)
    \cdot
    R'(\nu_n + \Delta\nu, \, \phi_n + \Delta\phi, \, \theta_n + \Delta\theta)
    }{
    \sum_{n=1}^{N_{\mathrm{s}}}
    M(\nu_n + \Delta\nu, \, \phi_n + \Delta\phi, \, \theta_n + \Delta\theta)
    },
    \label{eq:Stacking}
\end{equation}
where the sum is over the number of halos included in the stacking, $N_\mathrm{s}$, and $\nu_n$, $\phi_n$, and $\theta_n$ denote the nearest frequency, right ascension, and declination grid indices to halo \emph{n}. The mask $M(\nu_n + \Delta\nu, \, \phi_n + \Delta\phi, \, \theta_n + \Delta\theta)$ is 1 when the voxel lies within the observed frequency and declination ranges, and 0 otherwise. \\
To mitigate biases introduced by the stacking procedure, we additionally subtract a null stack, $r(\Delta \nu, \Delta \phi, \Delta \theta)$. This cube is constructed by applying the stacking procedure in eq. \eqref{eq:Stacking} to cutouts centred on random positions, with their frequency, right ascension, and declination coordinates drawn from the corresponding distributions of the halo catalogue. Since these random positions are not associated with halo locations, the resulting null stack is not expected to exhibit a centrally peaked signal. As described in section \ref{sec:Signal-to-Noise Ratio}, we forecast the stacking signal for different numbers of sources, $N_\mathrm{s}$, included in the stacking. However, since the null stack is not limited by the number of sources with known positions from an external survey, it is always constructed using the maximum number of positions considered in the analysis, such that $N_\mathrm{r} \geq N_\mathrm{s}$. In our analysis, we use $N_\mathrm{r} \simeq \num{1.15e8}$ random positions, corresponding to the number of halos within the observed sky region above a minimum halo mass of $10^{11} \, \mathrm{M}_{\odot}$; see also table \ref{tab:halo_counts}. The final bias-subtracted data cube is then given by:
\begin{equation}
    d(\Delta\nu, \Delta\phi, \Delta\theta) = s(\Delta\nu, \Delta\phi, \Delta\theta) - r(\Delta\nu, \Delta\phi, \Delta\theta).
    \label{eq:Data cube}
\end{equation}

\subsection{Signal-to-Noise Ratio}
\label{sec:Signal-to-Noise Ratio}
To estimate the signal-to-noise ratio, we fit a signal-only template cube to the data. The template is constructed by stacking HI-only ringmaps at the corresponding halo positions, following the procedure described in section \ref{sec:Stacked Spectral Cube}. As for the null stack, the template is constructed using the maximum number of halos considered in the analysis. The template cube, denoted by $t_i$, is fitted to the foreground-cleaned, noise-contaminated stacked cube, $d_i = s_i - r_i$, assuming a single free amplitude, $A$:
\begin{equation}
    d_i = A \cdot t_i + n_i,    
    \label{eq:template fit}
\end{equation}
where we have introduced an index notation in which \emph{i} runs over all voxels of the flattened cube.\footnote{$d_i = d(\Delta\nu, \Delta\phi, \Delta\theta)$.} The noise, $n_i$, is characterised by the covariance matrix $C_\mathrm{d}$, whose diagonal entries are given by the noise variance of each voxel, $\sigma_{n_i}^2$, while the off-diagonal entries account for correlations between different voxels. Assuming that the cross-covariance between the source and null stacks is negligible, the covariance of the final data cube is given by:
\begin{equation}
    C_\mathrm{d} = 
    C_\mathrm{s} + C_\mathrm{r} = 
    \frac{C}{N_\mathrm{s}} + \frac{C}{N_\mathrm{r}} = 
    \frac{C}{N_\mathrm{s}} \left(1 + \frac{N_\mathrm{s}}{N_\mathrm{r}} \right),
    \label{eq:Covariance matrix stack}
\end{equation}
where \emph{C} denotes the noise covariance of an individual cutout. We estimate \emph{C} from $N \simeq 10^6$ random cutouts, corresponding to the number of halos above a minimum halo mass of $10^{13} \, \mathrm{M}_{\odot}$; see table \ref{tab:halo_counts}:
\begin{equation}
    C_{ij} = 
    \frac{1}{N_{ij}-1} 
    \sum_{n=1}^N M_i^{(n)}M_j^{(n)} 
    \left(R{'}_i^{(n)} - r_i \right) 
    \left(R{'}_j^{(n)} - r_j \right), 
    \qquad
    N_{ij} = \sum_{n=1}^N M_i^{(n)} M_j^{(n)},
    \label{eq:Covariance matrix single cube}
\end{equation}
where $M_i^{(n)}$ is the mask associated with the \emph{n}th cutout and $R{'}_i^{(n)}$ denotes the corresponding random cutout.\footnote{$M_i^{(n)} = M(\nu_n + \Delta\nu, \, \phi_n + \Delta\phi, \, \theta_n + \Delta\theta)$ and $R{'}_i^{(n)} = R'(\nu_n + \Delta\nu, \, \phi_n + \Delta\phi, \, \theta_n + \Delta\theta)$.}
\\
The least-squares estimator for the template amplitude and its uncertainty are then given by:
\begin{equation}
    \hat{A} = 
    \frac{t^\mathsf{T} C_\mathrm{d}^{-1} d}{t^\mathsf{T} C_\mathrm{d}^{-1} t},
    \qquad
    \sigma_A = 
    \frac{1}{\left(t^\mathsf{T} C_\mathrm{d}^{-1} t \right)^{1/2}}
    ,
    \label{eq:Least-squares amplitude and uncertainty}
\end{equation}
yielding the template-detection signal-to-noise ratio
\begin{equation}
    \text{SNR} =
    \frac{\hat{A}}{\sigma_A} = \frac{t^\mathsf{T} C_\mathrm{d}^{-1} d}{\left(t^\mathsf{T} C_\mathrm{d}^{-1} t \right)^{1/2}}
    = 
    \frac{t^\mathsf{T} C^{-1} d}{\left(t^\mathsf{T} C^{-1} t \right)^{1/2}}
    \left( \frac{N_\mathrm{s}}{1 + N_\mathrm{s} / N_\mathrm{r}}\right)^{1/2}.
    \label{eq:SNR}
\end{equation} \\
The signal-to-noise ratio depends on several parameters. First, it scales approximately as $N_\mathrm{s}^{1/2}$ with the number of stacked halos.\footnote{For $N_\mathrm{r} \geq N_\mathrm{s}$, the contribution of the null stack to the total covariance is at most equal to that of the source stack.} The available number of sources is set by the survey used for the cross-correlation and increases approximately linearly with the overlapping sky area, $A_\mathrm{sky}$, such that $\mathrm{SNR} \propto A_\mathrm{sky}^{1/2}$ for a fixed source density. The source density itself depends on the survey's magnitude or flux limits and on additional selection criteria, which determine the completeness of the source sample. Second, the SNR depends on the observing time. As described in section \ref{sec:Instrument Model}, the instrumental noise standard deviation scales as $t_\mathrm{obs}^{-1/2}$. Consequently, $C^{-1} \propto t_\mathrm{obs}$ and hence $\mathrm{SNR} \propto t_\mathrm{obs}^{1/2}$. The SNR is further affected by the foreground filtering. Stronger filtering reduces foreground contamination but simultaneously removes part of the cosmological HI signal. As discussed in section \ref{sec:Visibilities}, the strength of the foreground filtering is parameterised by the delay cutoff, $\tau_\mathrm{cut}$. Finally, the SNR depends on the precision of the source redshifts. While the angular positions of the sources are typically well constrained, uncertainties in their redshifts translate into uncertainties in their radial positions. These uncertainties, characterised by the survey-dependent parameter $\sigma_z$, smear the stacked HI signal along the frequency axis and thereby reduce the SNR. \\
Based on these dependencies, we forecast the SNR as a function of the foreground delay cutoff, $\tau_\mathrm{cut}$, the redshift uncertainty, $\sigma_z$, and the number of stacked sources, $N_\mathrm{s}$, to determine the redshift precision and source sample size required from an external survey for a significant detection of the stacked HI signal under different foreground-filtering assumptions. Throughout this analysis, we fix the overlapping sky area and observing time to \qty{3547}{\deg^2} and 120 days, respectively. The redshift uncertainties are modelled by perturbing the true halo redshifts with random offsets drawn from a Gaussian distribution, $\mathcal{N}(0, \sigma_z^2)$. The number of sources available for stacking is controlled by two parameters: a minimum halo mass, $M_\mathrm{min}$, and a completeness fraction, $f_\mathrm{c}$. The minimum halo mass acts as a proxy for the magnitude or flux limit of an external survey and determines the underlying halo sample, while $f_\mathrm{c}$ accounts for additional selection effects by specifying the fraction of these halos that are retained in the stacking analysis. \\
Specifically, we evaluate the SNR for three representative redshift uncertainties, $\sigma_z / (1+z) = 0.001$, $0.01$, and $0.02$, evaluated at the central redshift $z=0.9$. These values are chosen to represent the expected performance of spectroscopic surveys (e.g. Euclid \cite{euclid_overview_2025}), optimised photometric samples (e.g. the DES redMaGiC galaxy catalogue \cite{DES_redMaGiC_2022}), and Stage IV photometric surveys (e.g. LSST \cite{LSST_2019}), respectively. To probe the dependence on the available source population, we vary the minimum halo mass over the range $10^{11} \, \mathrm{M}_{\odot} \leq M_\mathrm{min} \leq 10^{14} \,\mathrm{M}_{\odot}$ and consider completeness fractions of $f_\mathrm{c} = 1.0$, 0.5, and 0.1. Table \ref{tab:halo_counts} lists the values of $M_\mathrm{min}$ used in the SNR calculations and the corresponding total numbers of halos within the observed stripe. The choice of delay cutoffs, $\tau_\mathrm{cut}$, is discussed in the following section. 
\begin{table}
    \centering
    \begin{tabular}{
        S[table-format=1e2]
        S[table-format=9.0]
    }
        \toprule
        {$M_{\mathrm{min}}\,[\mathrm{M}_{\odot}]$} &
        {$N_{\mathrm{halos}}$} \\
        \midrule
        1e14 & 27434 \\
        5e13 & 99534 \\
        1e13 & 976433 \\
        5e12 & 2208666 \\
        1e12 & 12306770 \\
        5e11 & 24622401 \\
        1e11 & 115249387 \\
        \bottomrule
    \end{tabular}
    \caption{Number of halos in the observed HIRAX stripe above the different minimum halo-mass thresholds.}
    \label{tab:halo_counts}
\end{table} \\
For each combination of $\tau_\mathrm{cut}$, $\sigma_z$, $M_\mathrm{min}$, and $f_\mathrm{c}$, we generate a source stack, \emph{s}. The null stack, \emph{r}, and the signal template, \emph{t}, are generated separately for each combination of $\tau_\mathrm{cut}$ and $\sigma_z$, using the full source sample with $M_\mathrm{min}=10^{11} \, \mathrm{M}_{\odot}$ and $f_\mathrm{c}=1.0$. Similarly, the covariance matrix, \emph{C}, is estimated separately for each combination of $\tau_\mathrm{cut}$ and $\sigma_z$, but using $M_\mathrm{min}=10^{13} \, \mathrm{M}_{\odot}$ and $f_\mathrm{c}=1.0$. The resulting null stack, template, and covariance matrix can then be applied across the different source stacks.

\subsection{Delay Power Spectrum}
\label{sec:Delay Power Spectrum}
\begin{figure}
    \centering
    \includegraphics[width=1.0\linewidth]{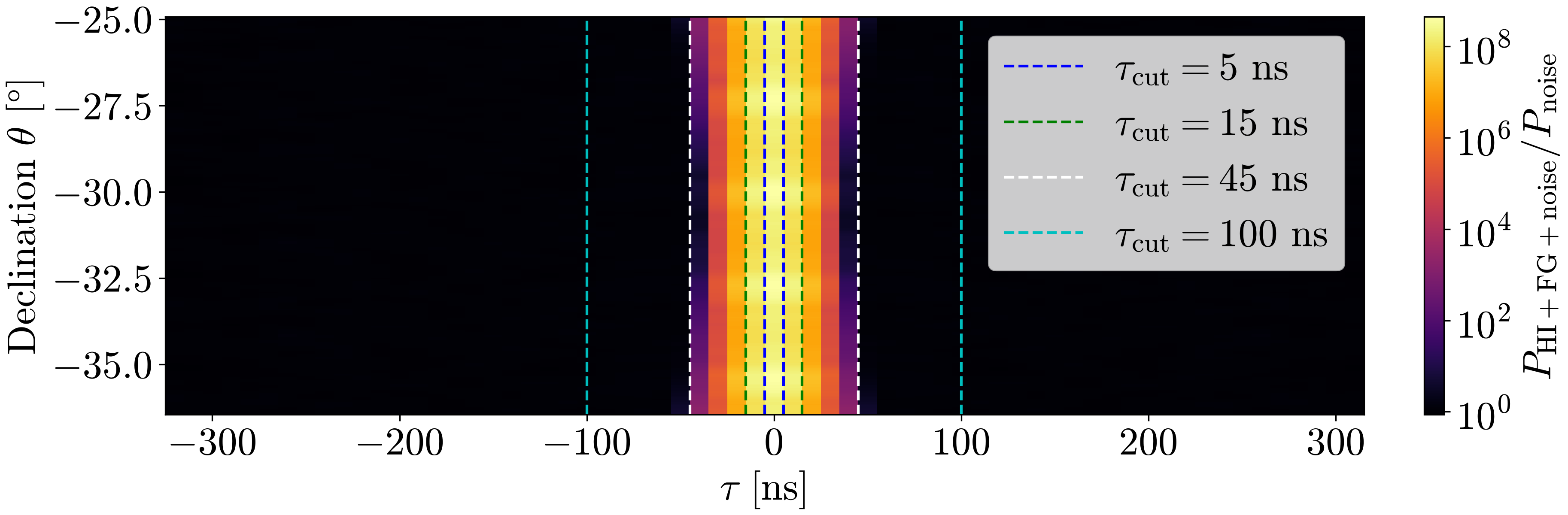}
    \caption{Delay power spectrum of the unfiltered ringmaps containing HI, foregrounds, and instrumental noise, normalised by the noise-only power spectrum, as a function of delay and declination. The dashed lines mark the four delay cutoffs adopted in this work to investigate the impact of progressively more aggressive foreground filtering on the stacking signal.}
    \label{fig:Delay Power Spectrum}
\end{figure}

To determine suitable values of $\tau_\mathrm{cut}$, we examine the delay power spectrum of the ringmaps in order to identify the delay modes that are most strongly contaminated by foreground emission. As in \cite{CHIME_stacking_eBOSS_2023}, the delay power spectrum is obtained by Fourier transforming the ringmaps along the frequency axis, yielding $\tilde{R}(\tau, \phi, \theta)$, and computing the variance over $\phi$,
\begin{equation}
    P(\tau, \theta) = \mathrm{Var}_\phi[\tilde{R}(\tau, \phi, \theta)].
    \label{eq:Delay Power Spectrum}
\end{equation}
Figure \ref{fig:Delay Power Spectrum} shows the delay power spectrum of the unfiltered ringmaps containing the HI signal, foregrounds, and instrumental noise, $P_\mathrm{HI+FG+noise}$, normalised by the corresponding noise-only power spectrum, $P_\mathrm{noise}$. As discussed in section \ref{sec:Visibilities}, the foreground contamination is concentrated at low delays. To investigate the impact of foreground filtering on the stacking signal, we consider four delay cutoffs in our analysis: $\tau_\mathrm{cut} = \qty{5}{\ns}$, \qty{15}{\ns}, \qty{45}{\ns}, and \qty{100}{\ns}. The two smallest cutoffs,  \qty{5}{\ns} and \qty{15}{\ns}, remove only the central delay modes, leaving significant foreground contamination in the data while preserving as much of the cosmological signal as possible. A cutoff of \qty{45}{\ns} removes all the foreground contamination in our simulations. As mentioned in section \ref{sec:Instrument Model}, we adopt a simplified beam model in this work, which is expected to underestimate the extent of foreground leakage in delay space. In practice, foreground emission is likely to contaminate substantially higher delays than those shown in figure \ref{fig:Delay Power Spectrum}, as a more realistic primary beam can introduce additional spectral structure. Therefore, the \qty{100}{\ns} cutoff represents a more realistic choice for HIRAX observations and allows us to assess the impact of such conservative foreground filtering on the resulting SNR. 
 \\
In this context, it is also useful to relate the Gaussian redshift uncertainties considered in our analysis to a characteristic delay scale. Using the redshift--frequency relation, $\nu = \nu_{21}/(1+z)$, a redshift uncertainty $\sigma_z$ corresponds to a frequency uncertainty $\sigma_\nu = \sigma_z \nu_{21}/(1+z)^2$. For Gaussian redshift uncertainties, the corresponding frequency smearing suppresses the signal in delay space by a factor of $\exp(-2\pi^2\sigma_\nu^2\tau^2)$, defining a characteristic delay scale $\tau_z=(2\pi\sigma_\nu)^{-1}$. At the central redshift of our simulation, $z=0.9$, the uncertainties $\sigma_z / (1+z) = 0.001$, $0.01$, and $0.02$ correspond to $\tau_z = \qty{213}{\ns}$, \qty{21}{\ns}, and \qty{11}{\ns}, respectively. Increasing the redshift uncertainty therefore progressively suppresses the stacking signal at higher delays. While for $\sigma_z/(1+z)=0.001$ substantial signal can remain beyond a realistic foreground cutoff of $\tau_\mathrm{cut}=\qty{100}{\ns}$, for $\sigma_z/(1+z)=0.01$ and $0.02$ the signal is already strongly suppressed at such delays. Consequently, a stacking detection in these cases may only be possible for the more optimistic foreground cutoffs of \qty{5}{\ns} and \qty{15}{\ns}.

\section{Results}
\label{sec:Results}
\subsection{SNR Analysis}
\label{sec:SNR Analysis}
\begin{figure}
    \centering
    \includegraphics[width=1\linewidth]{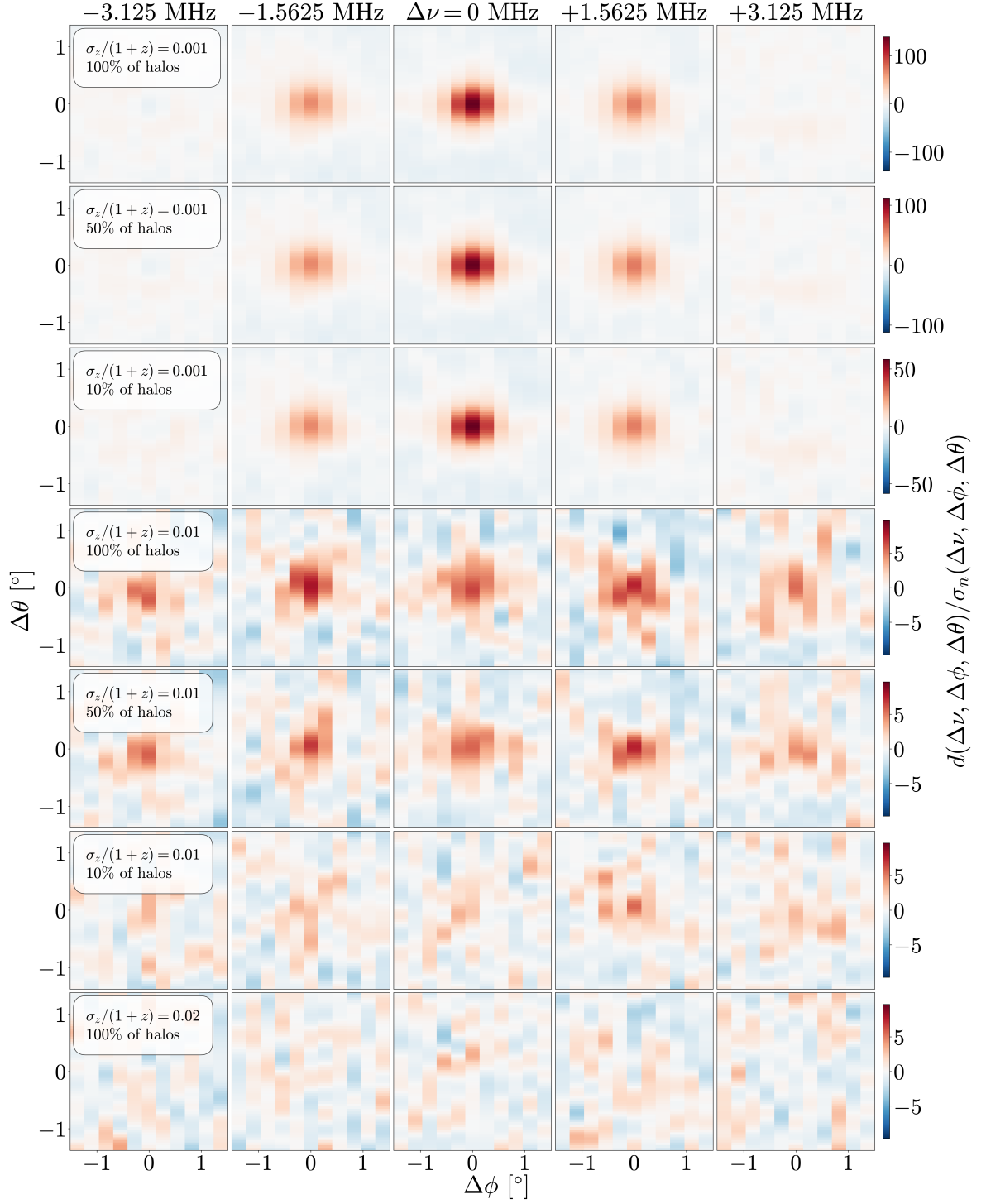}
    \caption{Bias-subtracted stacked spectral cubes obtained after applying a delay cutoff of $\tau_\mathrm{cut}=\qty{5}{\ns}$ to the ringmaps. The stacks include halos down to a minimum halo mass of $M_\mathrm{min}=10^{11} \, \mathrm{M}_{\odot}$. The stacked signal, $d(\Delta\nu, \Delta\phi, \Delta\theta)$, is normalised by the noise standard deviation in each voxel, $\sigma_n(\Delta\nu, \Delta\phi, \Delta\theta)$. Columns show slices at different frequency offsets, $\Delta\nu$, while rows correspond to different combinations of redshift uncertainty, $\sigma_z$, and halo completeness, $f_\mathrm{c}$.}
    \label{fig:Stacked signal tau=5ns}
\end{figure} 
\begin{figure}
    \centering
    \begin{subfigure}{0.49\linewidth}
        \centering
        \includegraphics[width=\linewidth]{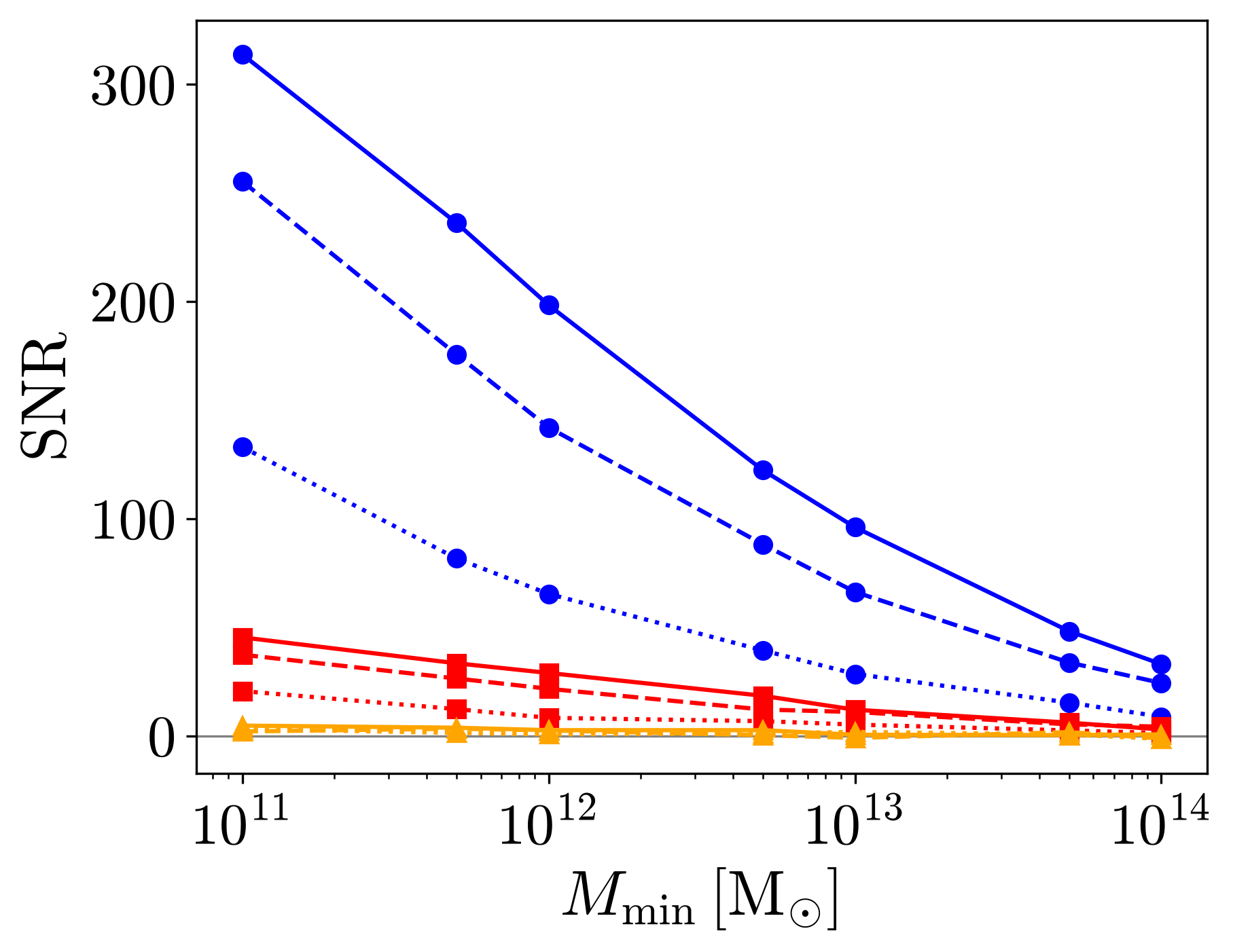}
        \caption{$\tau_\mathrm{cut} = \qty{5}{\ns}$}
    \end{subfigure}
    \hfill
    \begin{subfigure}{0.49\linewidth}
        \centering
        \includegraphics[width=\linewidth]{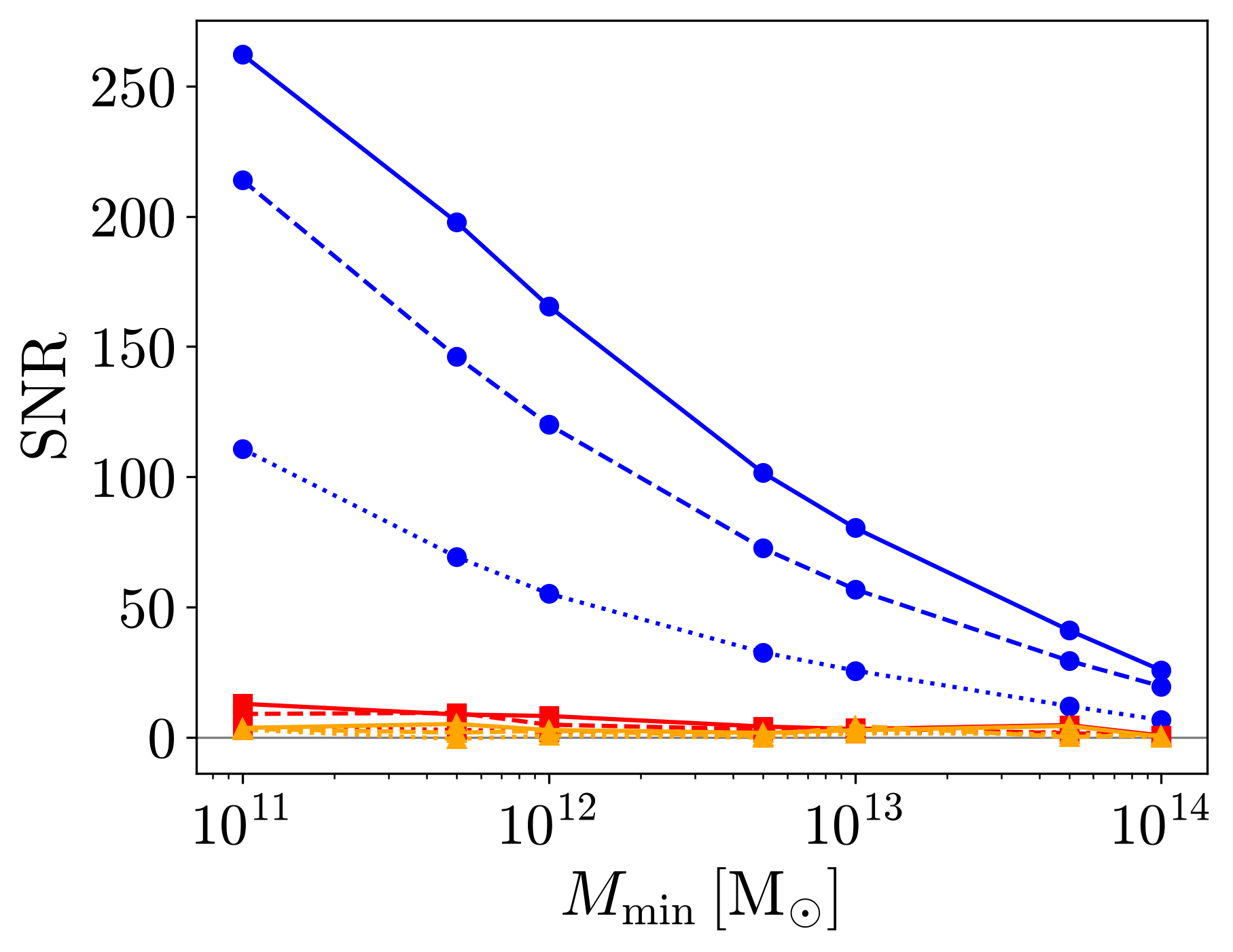}
        \caption{$\tau_\mathrm{cut} = \qty{15}{\ns}$}
    \end{subfigure}

    \vspace{0.5cm}

    \begin{subfigure}{0.49\linewidth}
        \centering
        \includegraphics[width=\linewidth]{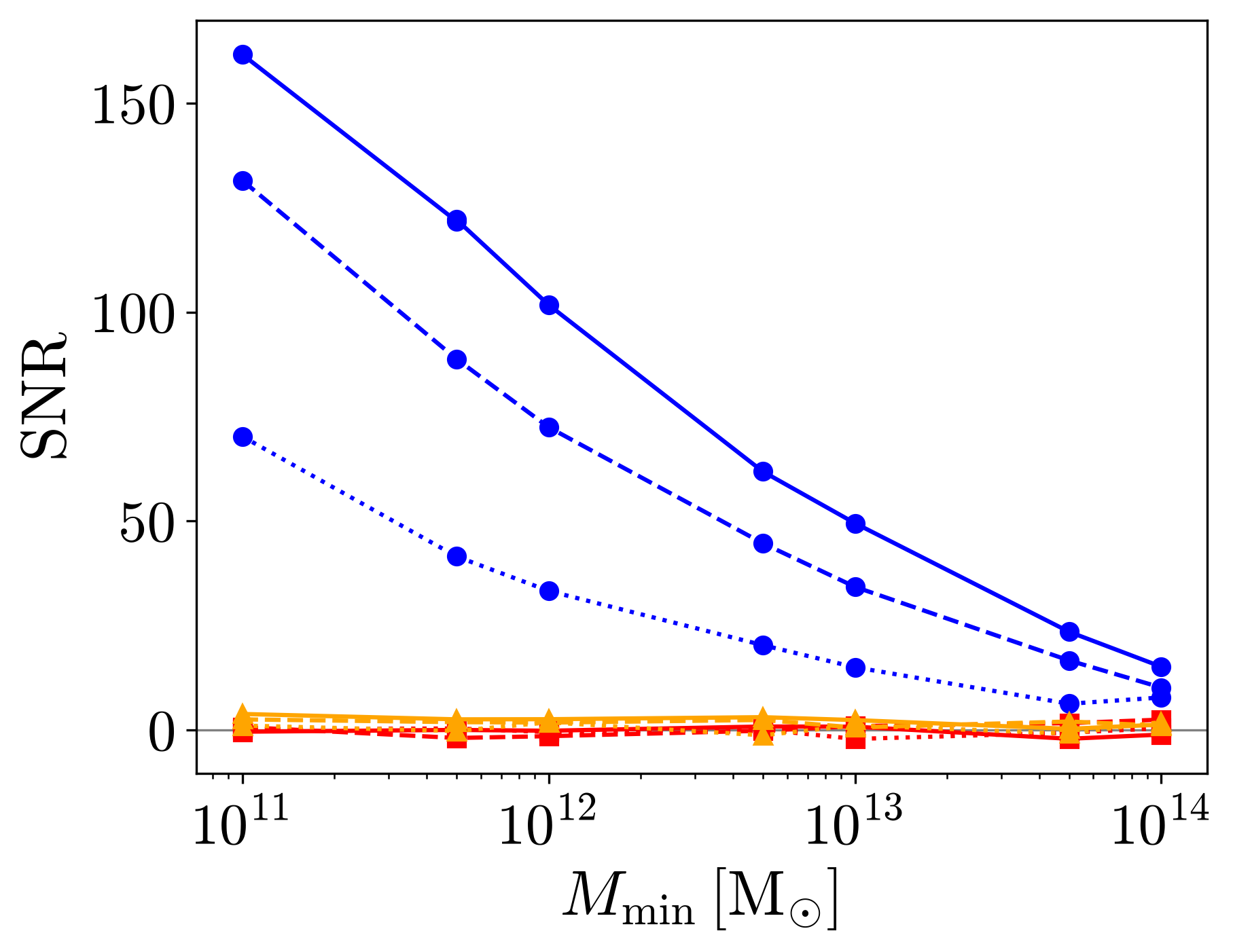}
        \caption{$\tau_\mathrm{cut} = \qty{45}{\ns}$}
    \end{subfigure}
    \hfill
    \begin{subfigure}{0.49\linewidth}
        \centering
        \includegraphics[width=\linewidth]{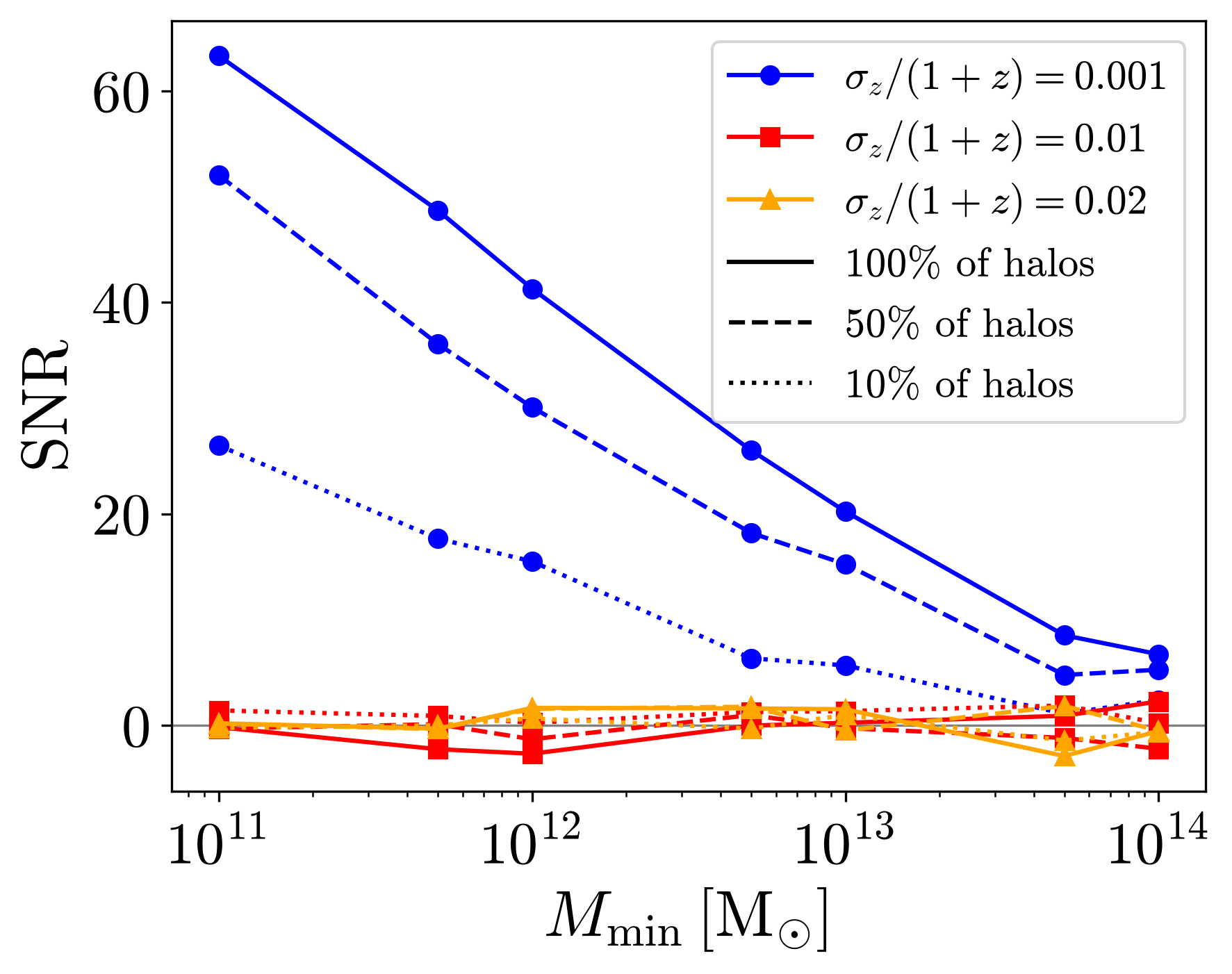}
        \caption{$\tau_\mathrm{cut} = \qty{100}{\ns}$}
    \end{subfigure}

    \caption{Signal-to-noise ratio (SNR), obtained from the template-fitting procedure described in section \ref{sec:Signal-to-Noise Ratio}, as a function of the minimum halo mass, $M_\mathrm{min}$, for different delay cutoffs, $\tau_\mathrm{cut}$. Each panel corresponds to a different value of $\tau_\mathrm{cut}$. The line colours and marker types indicate the assumed redshift uncertainty, $\sigma_z$, while the line styles denote the halo completeness fraction, $f_\mathrm{c}$. The SNR values assume a single HIRAX pointing at zenith, covering \qty{3547}{\deg^2}, and an observing time of 120 days. They can be rescaled to different survey areas and observing times according to $\mathrm{SNR} \propto A_\mathrm{sky}^{1/2} t_\mathrm{obs}^{1/2}$.}
    \label{fig:SNR}
\end{figure}
Figure \ref{fig:Stacked signal tau=5ns} shows the bias-subtracted stacked spectral cubes obtained after applying a delay cutoff of $\tau_\mathrm{cut} = \qty{5}{\ns}$ to the ringmaps and including all halos down to a minimum halo mass of $M_\mathrm{min} = 10^{11} \, \mathrm{M}_{\odot}$. For clarity, we show only the five central frequency channels, rather than all 21 channels. The signal is shown in units of the noise standard deviation in each voxel, $d(\Delta\nu, \Delta\phi, \Delta\theta) / \sigma_n(\Delta\nu, \Delta\phi, \Delta\theta)$. As described in section \ref{sec:Signal-to-Noise Ratio}, $\sigma_n^2(\Delta\nu, \Delta\phi, \Delta\theta)$ is given by the corresponding diagonal entries of the covariance matrix $C_\mathrm{d}$. Each column presents a slice through the stacked cube at a different frequency offset, $\Delta \nu$, relative to the centre, as indicated at the top of the figure. The horizontal and vertical axes show the right ascension and declination offsets from the cube centre, $\Delta \phi$ and $\Delta \theta$, respectively. Each row corresponds to a different combination of redshift uncertainty, $\sigma_z$, and completeness fraction, $f_\mathrm{c}$. The first three rows show the results for $\sigma_z / (1+z) = 0.001$ with progressively decreasing halo completeness, the next three rows correspond to $\sigma_z / (1+z) = 0.01$, and the last row shows $\sigma_z / (1+z) = 0.02$. For the latter case, only the full halo sample ($f_\mathrm{c}=1$) is shown, as no detectable signal remains even at full completeness, implying that lower completeness fractions would likewise yield no detectable signal. The corresponding stacked cubes for the remaining delay cutoffs are presented in appendix \ref{app:Stacked Spectral Cubes}. \\
As expected, when a stacking signal is present, it peaks at the centre of the cube, corresponding to $\Delta \nu = 0$, $\Delta \phi = 0$, and $\Delta \theta = 0$. The signal is strongest for the smallest redshift uncertainty and the largest number of stacked halos, and its amplitude decreases as the number of halos is reduced. Increasing the redshift uncertainty progressively broadens the signal along the frequency axis, causing it to spread over larger frequency offsets before ultimately disappearing as the correlation between the HI signal and the perturbed halo positions is lost. \\
Even from visual inspection, it is evident that for $\tau_\mathrm{cut} = \qty{5}{\ns}$, both $\sigma_z / (1+z) = 0.001$ and $\sigma_z / (1+z) = 0.01$ yield a stacking signal, even for reduced halo completeness. In contrast, a redshift uncertainty of $\sigma_z / (1+z) = 0.02$ completely suppresses the signal. The results presented in appendix \ref{app:Stacked Spectral Cubes} further show that for larger delay cutoffs, the signal is already lost for $\sigma_z / (1+z) = 0.01$. For spectroscopic redshifts, however, the stacking signal remains detectable even in the presence of catalogue incompleteness. As the delay cutoff increases, the signal amplitude gradually decreases, while the characteristic negative sidelobes along the frequency direction shift toward smaller frequency offsets. \\
The same trends are evident in figure \ref{fig:SNR}, which shows the SNR obtained from the template fitting as a function of the minimum halo mass, $M_\mathrm{min}$. The reported values correspond to a single HIRAX pointing at zenith, covering \qty{3547}{\deg^2}, and 120 days of observation. As discussed in section \ref{sec:Signal-to-Noise Ratio}, these results can be rescaled to different survey areas and observing times according to $\mathrm{SNR} \propto A_\mathrm{sky}^{1/2} t_\mathrm{obs}^{1/2}$. Each panel corresponds to a different value of $\tau_\mathrm{cut}$. The line colours and marker types denote different redshift uncertainties, while the line styles distinguish between different completeness fractions. As expected, the highest SNR is achieved for the smallest delay cutoff, the lowest redshift uncertainty, and the largest number of stacked halos. The SNR decreases systematically with increasing $\tau_\mathrm{cut}$ and $\sigma_z$, and with a decreasing number of halos, either due to a larger minimum halo mass, $M_\mathrm{min}$, or a lower completeness fraction, $f_\mathrm{c}$. \\
Our results demonstrate that the stacking signal is primarily limited by the redshift uncertainty. While spectroscopic redshifts preserve a significant signal even in the presence of substantial catalogue incompleteness and foreground filtering, photometric redshift errors progressively decorrelate the HI signal from the halo positions. Although a signal remains detectable for exceptionally good photometric redshifts, $\sigma_z/(1+z) = 0.01$, when a very low delay cutoff is applied, this disappears for realistic delay cutoffs. As discussed above, real data are expected to contain additional systematic effects with more complex delay structure than captured by our simulations. These effects would likely prevent a stacking detection using photometric redshifts at such a low delay cutoff. Moreover, even if such weak filtering were assumed, we could not identify any existing or planned photometric survey sample that combines sufficiently small redshift uncertainties with enough sources to enable a detection under these conditions. \\
We therefore conclude that, particularly for a realistic delay cutoff of $\tau_\mathrm{cut}=\qty{100}{\ns}$, a statistically significant HI detection through stacking is only feasible by cross-correlating HIRAX observations with overlapping surveys that provide sufficiently large samples with spectroscopic redshifts. Several large spectroscopic surveys, including DESI \cite{DESI_2016}, 4MOST \cite{4MOST_2019}, and Euclid \cite{euclid_overview_2025}, could provide suitable galaxy samples overlapping with the HIRAX survey footprint. Euclid is a particularly promising counterpart for an initial HIRAX zenith pointing. In the next section, we therefore forward model a Euclid-like observational scenario, adopting realistic assumptions for the redshift uncertainty and halo catalogue completeness, and present the resulting stacking signal.

\subsection{Euclid Stacking Forecasts}
\label{sec:Euclid Stacking Forecasts}
\begin{figure}
    \centering
    \includegraphics[width=0.75\linewidth]{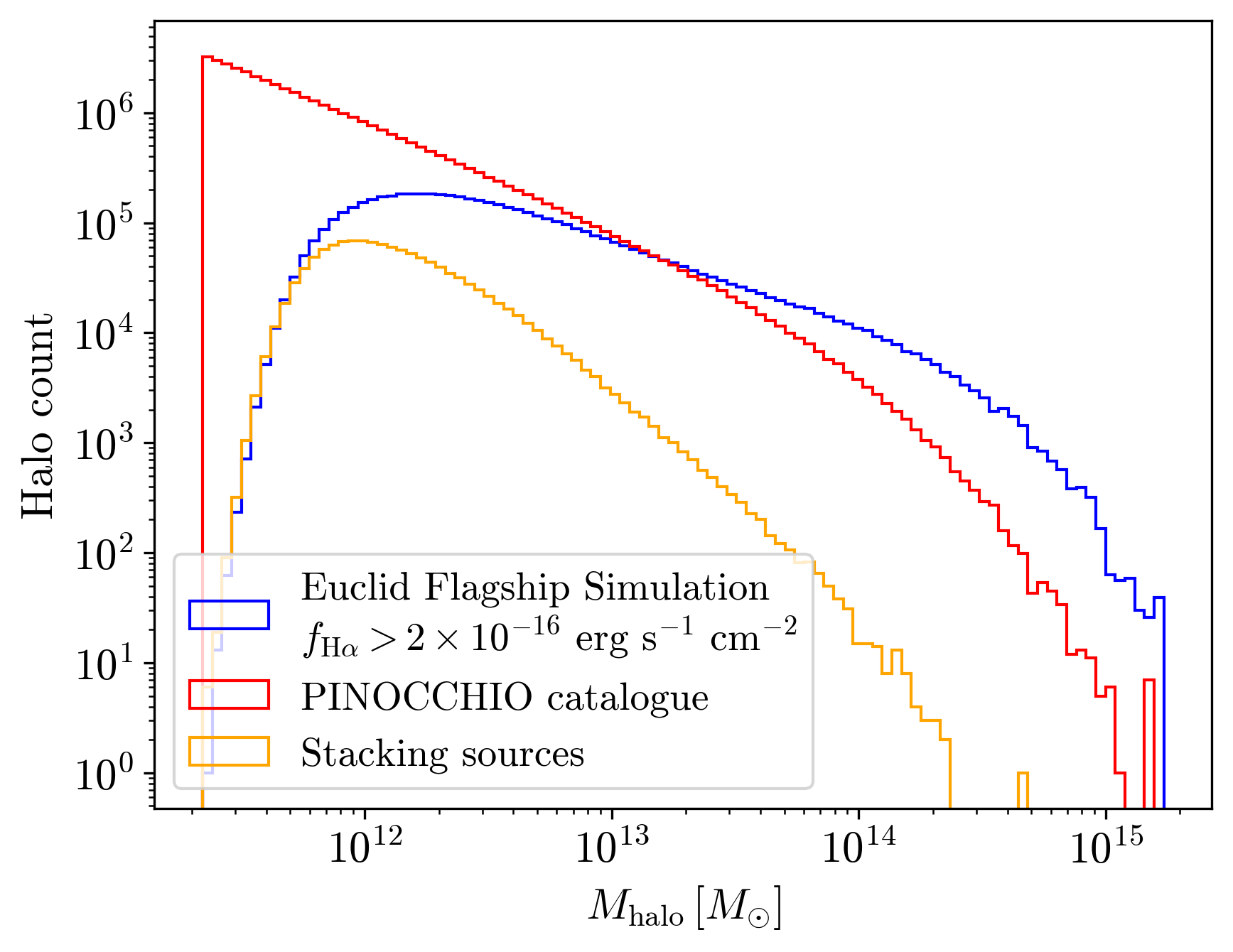}
    \caption{Halo mass distribution of the Euclid Flagship Simulation galaxy mock catalogue (blue) and the \texttt{PINOCCHIO} catalogue (red), both restricted to $0.77 < z < 1.03$. The Euclid Flagship sample includes halos within an octant of the sky that host H$\alpha$ emitters with fluxes $f_{\mathrm{H}\alpha} > \qty{2e-16}{erg \, \s^{-1} \cm^{-2}}$, while the \texttt{PINOCCHIO} sample includes halos with masses above \qty{2.2e11}{M_{\odot}} within the masked \qty{2562}{\deg^2} HIRAX survey stripe. The orange histogram shows the \num{1.18e6} halos randomly sampled from the \texttt{PINOCCHIO} sample using the normalised Euclid Flagship halo mass distribution as the sampling probability.}
    \label{fig:Euclid_samples}
\end{figure}
\begin{figure}
    \centering
    \includegraphics[width=1\linewidth]{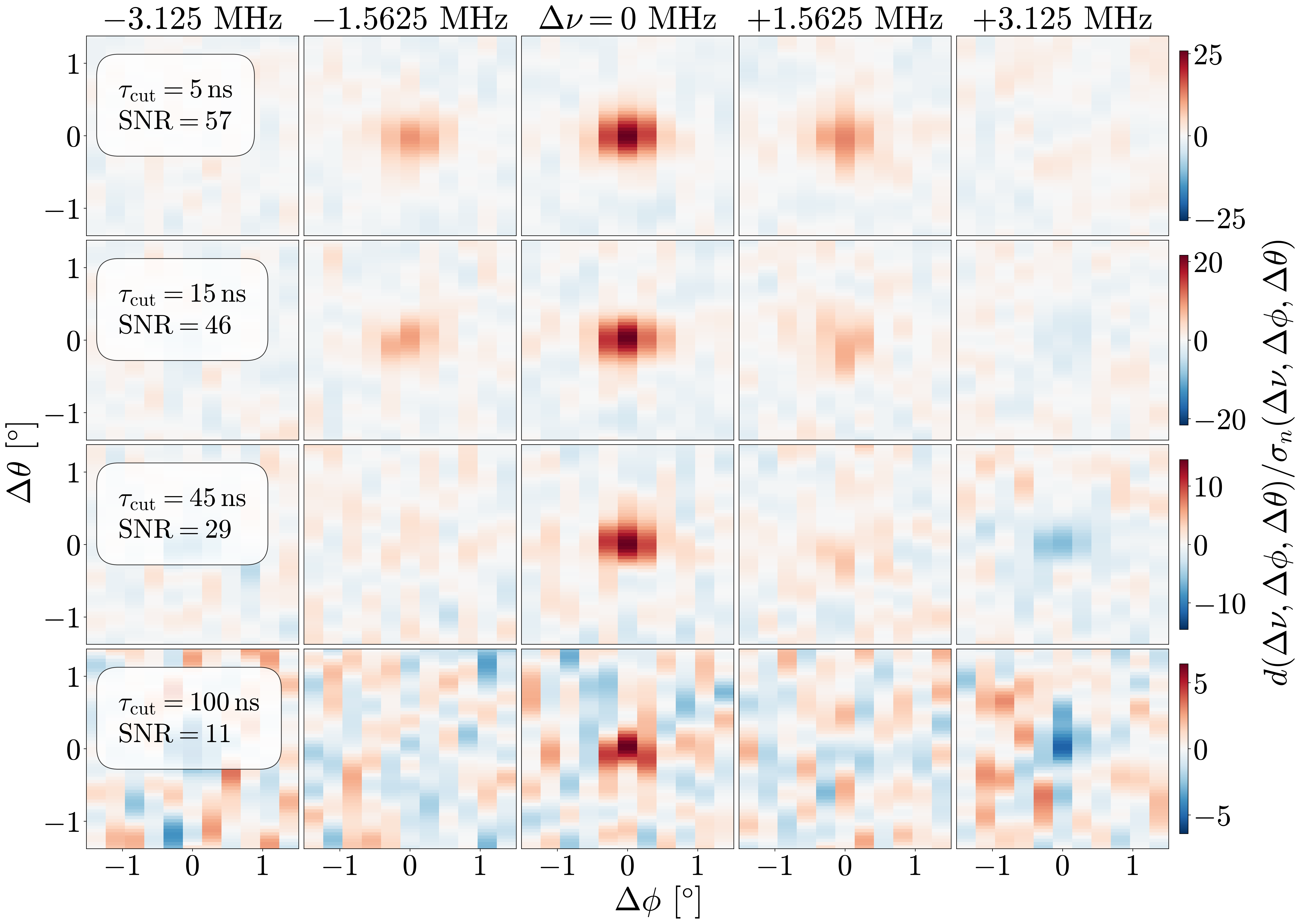}
    \caption{Bias-subtracted stacked spectral cubes for \num{1.18e6} Euclid-like halos sampled from the \texttt{PINOCCHIO} catalogue. The sample follows the host-halo mass distribution of Euclid Flagship H$\alpha$ emitters with $f_{\mathrm{H}\alpha} > \qty{2e-16}{erg \, \s^{-1} \cm^{-2}}$, with a minimum halo mass of $M_\mathrm{min}=\qty{2.2e11}{M_{\odot}}$. Gaussian redshift uncertainties of $\sigma_z/(1+z)=0.001$ are assumed. The stacked signal, $d(\Delta\nu, \Delta\phi, \Delta\theta)$, is normalised by the noise standard deviation in each voxel, $\sigma_n(\Delta\nu, \Delta\phi, \Delta\theta)$. Each row corresponds to a different delay cutoff, $\tau_\mathrm{cut}$, with the resulting SNR from template fitting indicated in each text box.}
    \label{fig:Euclid stacked signal} 
\end{figure}
\begin{figure}
    \centering
    \includegraphics[width=0.75\linewidth]{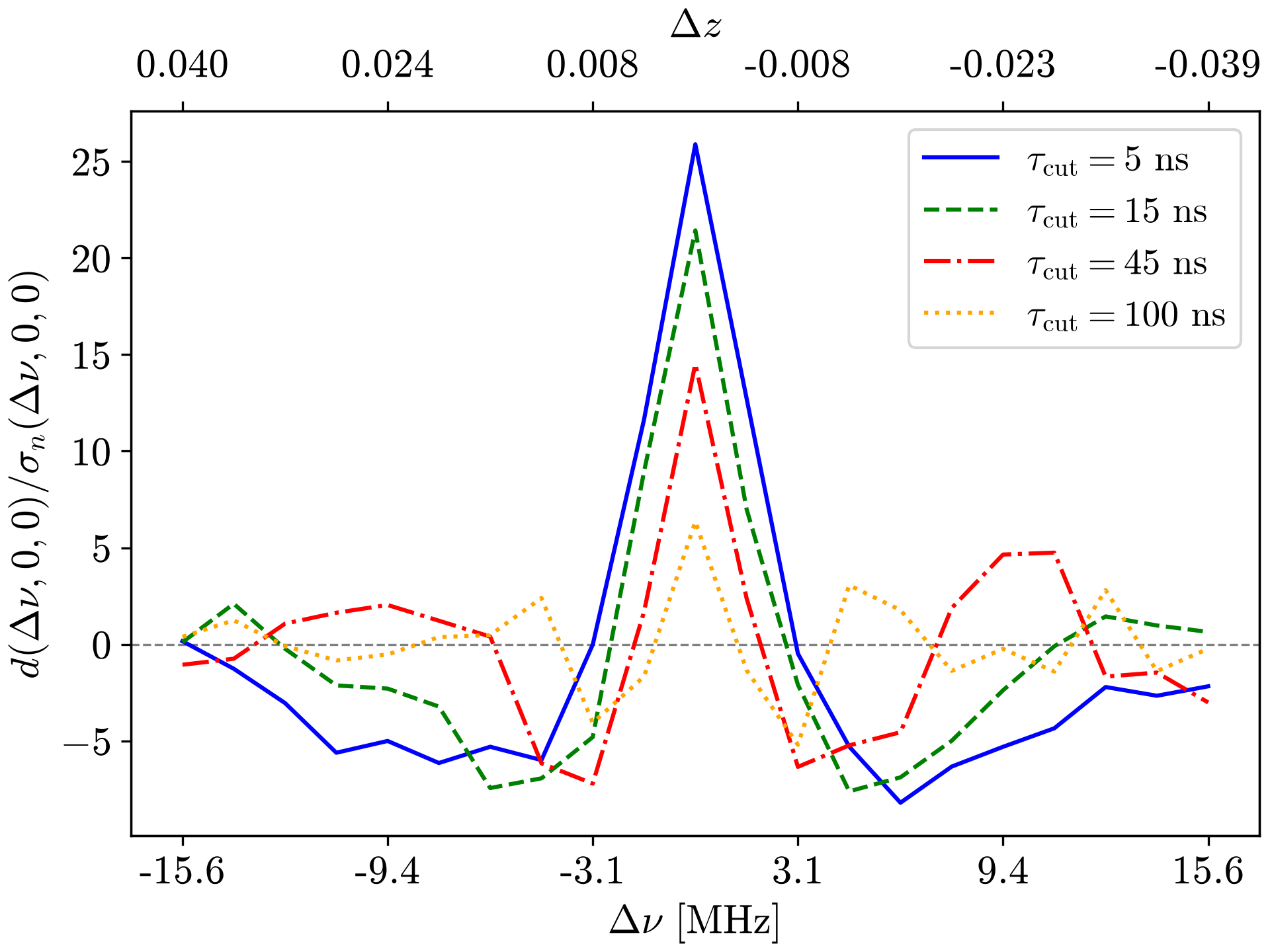}
    \caption{Bias-subtracted stacked HI signal in the central spatial pixel, $\Delta \phi = \qty{0}{\degree}$ and $\Delta \theta = \qty{0}{\degree}$, of the Euclid-stacking cube shown in figure \ref{fig:Euclid stacked signal}, as a function of the frequency offset, $\Delta\nu$, from the centre of the cube. The signal is normalised by the noise standard deviation in the corresponding voxel. The different lines correspond to delay cutoffs of $\tau_\mathrm{cut} = \qty{5}{\ns}$, $\qty{15}{\ns}$, $\qty{45}{\ns}$, and $\qty{100}{\ns}$. The upper x-axis indicates the redshift offset corresponding to each frequency offset.}
    \label{fig:Euclid stacked signal freq direction}
\end{figure}
Euclid will conduct two major surveys: the Euclid Wide Survey, covering approximately \qty{14000}{\deg^2} of the sky, and the Euclid Deep Survey, covering approximately \qty{53}{\deg^2} \cite{euclid_overview_2025}. As our analysis requires a large sky area, we focus exclusively on the Euclid Wide Survey. Its two principal science objectives are the measurement of weak gravitational lensing using photometric galaxy shape measurements and galaxy clustering from spectroscopic observations. Here, we exploit the high redshift precision of the spectroscopic galaxy sample to maximise the signal-to-noise ratio of the stacked HI signal. \\
Euclid uses slitless spectroscopy to measure the redshifts of H$\alpha$ emitters over the redshift range $z=0.9$--1.8 with an uncertainty of $\sigma_z/(1+z) < 0.001$ \cite{euclid_overview_2025}.\footnote{The instrument is expected to detect H$\alpha$ over the broader redshift range $z=0.84$--1.88. However, the mission requirements are typically quoted for the more conservative range.} For the redshift interval $z=0.90$--1.10, the expected number density is $dN(z_\mathrm{mean})/d\Omega dz = \qty{1815}{deg^{-2}}$ galaxies per unit area and redshift interval \cite{euclid_preparation_2020}. For simplicity, we assume that this density can be extrapolated to our simulated HIRAX redshift range of $z=0.77$--1.03, as we are primarily interested in obtaining a representative total sample size for the stacking analysis. A dedicated simulation covering the Euclid redshift range would exclude sources at $z<0.9$, but this reduction in sample size could be compensated by including additional sources at higher redshifts $z>1.03$. We therefore do not expect this to significantly affect the forecast presented here. Additionally, we mask the Galactic plane by excluding $\pm \qty{25}{\degree}$ around $\phi=\qty{120}{\degree}$ and \qty{266}{\degree}. This reduces the sky area of the simulated HIRAX stripe from \qty{3547}{\deg^2} to \qty{2562}{\deg^2}. Under these assumptions, we expect \num{1.18e6} Euclid spectroscopic sources within the overlapping survey area, which can be used in the HI stacking analysis. \\
Euclid is expected to detect H$\alpha$ emitters with line fluxes \qty{>2e-16}{erg \, \s^{-1} \cm^{-2}} at \qty{1600}{\nm} \cite{euclid_overview_2025}. To relate this flux limit to a minimum halo mass as a selection criterion for the \texttt{PINOCCHIO} halos, we make use of the Euclid Flagship Simulation galaxy mock catalogue \cite{euclid_flagship_simulation_2025}. The underlying dark matter simulation was performed with the \texttt{PKDGRAV3} \cite{PKDGRAV3_2017} \emph{N}-body code. It consists of a full-sky particle light cone extending to $z=3$, generated by replicating a simulation box of side length \qty{3600}{\emph{h}^{-1} Mpc} containing $\num{16000}^3$ particles.\footnote{The Euclid Flagship Simulation adopts $h=0.67$.} Halos were identified using the \texttt{ROCKSTAR} halo finder \cite{ROCKSTAR_2013}, requiring a minimum of 10 particles per halo. This yields approximately \num{126e+9} main halos with a minimum halo mass of \qty{1.5e+10}{\mathrm{M}_{\odot}}. Galaxies were subsequently assigned to the halos in a single octant of the light cone, corresponding to approximately \num{15.8e+9} halos, using a combination of halo occupation distribution and abundance matching. Using the dust-extinguished H$\alpha$ fluxes calibrated to model 3 of \cite{Pozzetti_2016}, we find that the least massive halos hosting galaxies that satisfy the Euclid flux limit have masses of \qty{2.2e11}{M_{\odot}}. For the halo mass, we use the bound halo mass reported in the Flagship catalogue, defined as the sum of the masses of all bound particles within the virial radius.\footnote{This is also the default halo mass definition adopted in \cite{euclid_flagship_simulation_2025}.} \\
The blue histogram in figure \ref{fig:Euclid_samples} shows the halo mass distribution of halos that host H$\alpha$ emitters above the Euclid flux limit in the redshift range $0.77 < z < 1.03$ for the simulated Flagship octant. The red histogram shows the mass distribution of \texttt{PINOCCHIO} halos with masses \qty{>2.2e11}{M_{\odot}} in the masked \qty{2562}{\deg^2} HIRAX survey stripe. To select the stacking sources, we randomly draw \num{1.18e6} halos from the masked \texttt{PINOCCHIO} catalogue with masses \qty{>2.2e11}{M_{\odot}}, using the normalised Flagship halo mass distribution as the sampling probability. This corresponds to approximately 2\% of all \texttt{PINOCCHIO} halos above the mass threshold within the unmasked simulated \qty{3547}{\deg^2} HIRAX survey stripe. The resulting mass distribution of the sampled halos is shown by the orange histogram. \\
Figure \ref{fig:Euclid stacked signal} shows the bias-subtracted stacked HI signal obtained from the \num{1.18e6} halos, assuming Gaussian redshift uncertainties of $\sigma_z/(1+z)=0.001$ at $z=0.9$. As before, we show only the five central frequency channels and express the signal in units of the noise standard deviation in each voxel, $d(\Delta\nu, \Delta\phi, \Delta\theta) / \sigma_n(\Delta\nu, \Delta\phi, \Delta\theta)$. Each row now corresponds to a different delay cutoff. Figure \ref{fig:Euclid stacked signal freq direction} shows the signal in the central spatial pixel, $\Delta \phi = \qty{0}{\degree}$ and $\Delta \theta = \qty{0}{\degree}$, as a function of the frequency offset, $\Delta \nu$, from the centre of the cube, with the different lines corresponding to the different delay cutoffs. \\
Both figures show a pronounced HI signal peak at the centre of the cube, $\Delta \nu = \qty{0}{\MHz}$, for all delay cutoffs. They also clearly illustrate the progressive suppression of the signal with increasingly aggressive foreground filtering. From the template fitting, we obtain signal-to-noise ratios of 57, 46, 29, and 11 for $\tau_\mathrm{cut} = $ \qty{5}{\ns}, \qty{15}{\ns}, \qty{45}{\ns}, and \qty{100}{\ns}, respectively. Thus, even for the most conservative cutoff of $\tau_\mathrm{cut} = \qty{100}{\ns}$, a significant stacking signal remains. This demonstrates that stacking HIRAX observations at the positions of Euclid spectroscopic galaxies provides a promising strategy for achieving a first detection of the HI signal. \\
Both the instrumental model and the noise estimate adopted here are optimistic: residual foreground contamination and instrumental systematics are expected to further reduce the detection significance in real observations. However, the present analysis is also restricted to a single HIRAX pointing and does not exploit the full HIRAX redshift range. Extending the analysis to a larger survey volume through additional pointings or a broader redshift range would increase the number of sources available for stacking and could therefore improve the detection significance.

\section{Conclusion}
\label{sec:Conclusion}
In the post-reionization Universe, most HI resides in self-shielded regions within galaxies and therefore traces the underlying matter distribution. Consequently, \qty{21}{cm} intensity mapping provides a promising probe of the three-dimensional large-scale structure of the Universe over vast cosmological volumes, complementary to traditional galaxy surveys. Instead of detecting and resolving individual galaxies, it measures the integrated emission from unresolved HI, while the well-defined rest-frame frequency of the hyperfine \qty{21}{\cm} transition allows redshift information to be extracted directly from the observed frequency of the emission. Leveraging this potential, however, requires extracting the very faint cosmological signal in the presence of astrophysical foreground emission that is orders of magnitude brighter and instrumental noise. \\
In this work, we have presented a forward modelling framework for forecasting statistical detections of the post-reionization cosmological HI signal with HIRAX through cross-correlation with external galaxy surveys. We have focused specifically on source stacking as a cross-correlation technique and investigated its prospects for enabling an early HI detection during the initial stages of HIRAX deployment. Starting from the fast halo-model-based HI simulations developed and validated in our previous work \cite{Hitz_2025}, we propagated the cosmological signal, together with astrophysical foreground emission from Galactic synchrotron emission and continuum radio point sources, as well as instrumental noise, through a model of a preliminary HIRAX array. Based on the \emph{m}-mode formalism \cite{Shaw_2014, Shaw_2015} and using the Python packages \texttt{cora}, \texttt{caput}, \texttt{driftscan}, and \texttt{draco} \cite{cora_2024, caput_2024, driftscan_2026, draco_2024}, we generated simulated visibilities, applied delay-domain foreground filtering with the \texttt{DAYENU} technique \cite{DAYENU_2021}, and constructed ringmaps following the map-making approach developed for CHIME \cite{CHIME_2014}. We then used these maps to investigate the detection of the HI signal through stacking at the three-dimensional positions of known sources. \\
To model an early stage of the HIRAX telescope, we considered a $6\times6$ array operating in drift-scan mode for 120 days with a fixed pointing at zenith. We restricted the frequency range to 700--\qty{800}{\MHz}, matching the coverage of the HI simulation presented in \cite{Hitz_2025}. Over a full rotation of the Earth, the observation covers a stripe of approximately \qty{3547}{\deg^2} extending $\pm \qty{5.7}{\degree}$ in declination around zenith. Despite the comparatively small array configuration and limited survey volume, our simulations show that the cosmological HI signal can be recovered statistically through source stacking, provided that a sufficiently large external source catalogue is available with precise individual redshifts. We quantified the requirements on such a catalogue by varying its redshift uncertainty ($\sigma_z/(1+z) = 0.001, 0.01$, and 0.02), number of available sources ($\sim \qtyrange{2.7e4}{1.15e8}{}$), and completeness (100\%, 50\%, and 10\%), while simultaneously investigating the impact of increasingly aggressive foreground filtering ($\tau_\mathrm{cut} = $ \qty{5}{\ns}, \qty{15}{\ns}, \qty{45}{\ns}, and \qty{100}{\ns}). \\
The resulting signal-to-noise ratio depends strongly on both the foreground filtering and the properties of the source catalogue. Increasing the delay cutoff removes an increasing fraction of the cosmological signal together with the foreground-contaminated modes and therefore reduces the achievable signal-to-noise ratio. More importantly, we find that the precision of the source redshifts is the most critical requirement for stacking. Redshift uncertainties broaden the stacked signal along the frequency direction and eventually decorrelate the source positions from the measured HI emission. While redshift uncertainties of $\sigma_z/(1+z)=0.01$ can still permit a detection for very weak foreground filtering, $\tau_\mathrm{cut} = \qty{5}{\ns}$, the signal is lost once more conservative delay cuts are applied. For the more realistic value of $\tau_\mathrm{cut}=\qty{100}{\ns}$ adopted for HIRAX, which serves as a proxy for the increased foreground leakage expected from a more complex beam model, a significant detection therefore requires spectroscopic redshift precision, $\sigma_z / (1+z) \leq 0.001$. In contrast, catalogue incompleteness primarily reduces the signal-to-noise ratio through the smaller number of stacking sources, and a detectable signal can be retained even for substantially incomplete catalogues when spectroscopic redshifts are available. \\
Motivated by these findings, we considered the future Euclid spectroscopic galaxy sample \cite{euclid_overview_2025} as a potential external source catalogue for HIRAX. Using the expected number density and redshift precision of Euclid H$\alpha$ emitters, together with the Euclid Flagship Simulation to relate the H$\alpha$ flux selection to the underlying halo population, we constructed a Euclid-like source sample containing \num{1.18e6} objects within the simulated HIRAX survey region. For this sample, the stacked HI signal remains clearly detectable for all foreground-filtering scenarios considered. Even for the most conservative cutoff of $\tau_\mathrm{cut} = \qty{100}{\mathrm{ns}}$, we obtain a template-fitting signal-to-noise ratio of approximately 11. This demonstrates that the combination of HIRAX intensity maps with an extensive spectroscopic galaxy sample provides a promising route towards an early detection of the post-reionization HI signal, and we anticipate this will be possible before the completion of the full HIRAX array. However, due to the limited availability of suitable spectroscopic surveys in the southern hemisphere and given the uncertainty in the release timeline of the Euclid spectroscopic data, it may also be preferable to initially target a non-zenith HIRAX pointing at declinations closer to the celestial equator, providing overlap with the DESI survey.\\
The forecasts presented here should be interpreted in the context of several simplifying assumptions. In particular, we model the instrumental response as an idealised Gaussian primary beam and neglect beam calibration errors and uncertainties, radio-frequency interference (RFI), polarisation leakage, and other instrumental systematics expected in real observations. Such effects can introduce additional spectral structure and increase foreground contamination. Although these effects are not included in our simulations, we have made methodological choices aimed at improving the robustness of our analysis pipeline to some of them. These include the use of \texttt{DAYENU} foreground filtering, which remains effective in the presence of RFI flagging, and the use of ringmaps as a minimally processed data product. Nevertheless, the resulting signal-to-noise ratios should be interpreted as forecasts under idealised assumptions rather than precise predictions of the detection significance achievable with real observations.\\
At the same time, the present setup does not exploit the full sensitivity and survey volume ultimately available to HIRAX. Our instrument model includes only 36 dishes, compared to the 1024 dishes planned for the completed array, and we consider only a single pointing and a limited redshift interval. The completed HIRAX survey is planned to cover a substantially larger fraction of the southern sky and probe the HI signal out to $z\sim2.5$. Extending the analysis to more dishes, additional pointings, and a wider redshift range will further improve the signal-to-noise ratio of the measurement. \\
The forward modelling pipeline developed in this work provides a flexible framework for connecting cosmological HI simulations directly to realistic interferometric data products. In future work, it can be extended to incorporate more realistic descriptions of the HIRAX instrument and its systematics, allowing their effects to be propagated through the entire observation and analysis pipeline. As HIRAX progresses through commissioning and successive deployment stages, such simulations can play an important role in developing and validating more advanced strategies for detecting the cosmological HI signal from early observations and in preparing the analysis methods required for the full survey.

\acknowledgments
This work was supported in part by grant 200021\_192243 from the Swiss National Science Foundation and by SERI as part of the SKACH consortium. \\
This work has made use of the \texttt{NumPy} \cite{numpy}, \texttt{SciPy} \cite{scipy}, \texttt{Matplotlib} \cite{matplotlib}, \texttt{HEALPix}/\texttt{healpy} \cite{HEALPix_2005, healpy}, \texttt{h5py} \cite{h5py}, and \texttt{Zarr} \cite{zarr} software packages. \\
We also made use of CosmoHub \cite{CosmoHub_1, CosmoHub_2}, developed by PIC (maintained by IFAE and CIEMAT) in collaboration with ICE-CSIC. It received funding from the Spanish government (grant EQC2021-007479-P funded by MCIN/AEI/10.13039/501100011033), the EU NextGeneration/PRTR (PRTR-C17.I1), and the Generalitat de Catalunya.

\appendix
\section{Stacked Spectral Cubes}
\label{app:Stacked Spectral Cubes}
Figures \ref{fig:Stacked signal tau=15ns}, \ref{fig:Stacked signal tau=45ns}, and \ref{fig:Stacked signal tau=100ns} show the bias-subtracted noise-normalised stacked spectral cubes for the remaining delay cutoffs, $\tau_\mathrm{cut} = $ \qty{15}{\ns}, \qty{45}{\ns}, and \qty{100}{\ns}, which were not included in the main text (section \ref{sec:SNR Analysis}). These results are also obtained by including all halos down to a minimum halo mass of $M_\mathrm{min} = 10^{11} \, \mathrm{M}_{\odot}$.

\begin{figure}
    \centering
    \includegraphics[width=1\linewidth]{figures/stacked_spectral_cubes_halo_minus_random_mmin1e11_dayenufiltered0.015.png}
    \caption{Same as figure \ref{fig:Stacked signal tau=5ns}, but for $\tau_\mathrm{cut} = \qty{15}{\ns}$.}
    \label{fig:Stacked signal tau=15ns}
\end{figure}

\begin{figure}
    \centering
    \includegraphics[width=1\linewidth]{figures/stacked_spectral_cubes_halo_minus_random_mmin1e11_dayenufiltered0.045.png}
    \caption{Same as figure \ref{fig:Stacked signal tau=5ns}, but for $\tau_\mathrm{cut} = \qty{45}{\ns}$.}
    \label{fig:Stacked signal tau=45ns}
\end{figure}

\begin{figure}
    \centering
    \includegraphics[width=1\linewidth]{figures/stacked_spectral_cubes_halo_minus_random_mmin1e11_dayenufiltered0.1.png}
    \caption{Same as figure \ref{fig:Stacked signal tau=5ns}, but for $\tau_\mathrm{cut} = \qty{100}{\ns}$.}
    \label{fig:Stacked signal tau=100ns}
\end{figure}

\newpage

\bibliographystyle{JHEP}
\bibliography{biblio.bib}

\end{document}